# Few-view single photon emission computed tomography (SPECT) reconstruction based on a blurred piecewise constant object model

**Paul A Wolf[1], Jakob H Jørgensen[2], Taly G Schmidt[1] and Emil Y Sidky[3]**

[1] Department of Biomedical Engineering, Marquette University, 1515 W. Wisconsin Ave. Milwaukee, WI 53233, USA

[2] Department of Informatics and Mathematical Modeling, Technical University of Denmark, Richard Petersens Plads, Building 321, 2800 Kgs. Lyngby, Denmark

[3] Department of Radiology, University of Chicago, 5841 S. Maryland Ave., Chicago, IL 60637, USA

Email: paul.wolf@marquette.edu, jakj@imm.dtu.dk, tal.gilat-schmidt@marquette.edu and sidky@uchicago.edu

**Short title.** Few-view SPECT reconstruction based on a blurred piecewise constant object model

**Abstract.** A sparsity-exploiting algorithm intended for few-view Single Photon Emission Computed Tomography (SPECT) reconstruction is proposed and characterized. The algorithm models the object as piecewise constant subject to a blurring operation. To validate that the algorithm closely approximates the true object in the noiseless case, projection data were generated from an object assuming this model and using the system matrix. Monte Carlo simulations were performed to provide more realistic data of a phantom with varying smoothness across the field of view. Reconstructions were performed across a sweep of two primary design parameters. The results demonstrate that the algorithm recovers the object in a noiseless simulation case. While the algorithm assumes a specific blurring model, the results suggest that the algorithm may provide high reconstruction accuracy even when the object does not match the assumed blurring model. Generally, increased values of the blurring parameter and TV weighting parameters reduced noise and streaking artifacts, while decreasing spatial resolution. As the number of views decreased from 60 to 9 the accuracy of images reconstructed using the proposed algorithm varied by less than 3%. Overall, the results demonstrate preliminary feasibility of a sparsity-exploiting reconstruction algorithm which may be beneficial for few-view SPECT.

## 1. Introduction

Single Photon Emission Computed Tomography (SPECT) provides noninvasive images of the distribution of radiotracer molecules. Dynamic Single Photon Emission Computed Tomography provides information about tracer uptake and washout from a series of time-sequence images. Dynamic SPECT acquisition methods measuring time activity curves on the order of minutes have been developed (Gullberg *et al* 2010, Gullberg 2004). However, the dynamic wash-in wash-out of some tracers occurs over a period of just several seconds, requiring better temporal sampling. Stationary ring-like multi-camera systems are being developed to provide rapid dynamic acquisitions with high temporal sampling (Beekman *et al* 2005, Furenlid *et al* 2004, Beekman and Vastenhouw 2004). Reducing the number of cameras reduces the cost of such systems but also reduces the number of views acquired, limiting the angular sampling of the system. Novel few-view image reconstruction methods may be beneficial and are being investigated for the application of dynamic SPECT (Ma *et al* 2012).

The feasibility of reconstructing from angularly undersampled, or few-view data, has recently been explored for CT (Sidky and Pan 2008, Chen et al 2008, Duan et al 2009, Ritschl et al 2011, Sidky et al 2006). These investigations are based on exploitation of gradient-magnitude sparsity, an idea promoted and theoretically investigated in the field of Compressed Sensing (CS). Few-view, sparsity-exploiting CT reconstruction algorithms promote gradient-magnitude sparsity by minimizing image total variation (TV). Success of these algorithms in allowing sampling reduction follows from an object model which is approximately piecewise constant, a model that may not apply well for SPECT objects. The SPECT object function quantifies the physiological uptake of a radiolabelled tracer in the body. In some applications, the transition between different uptake regions in the SPECT object is expected to be smoother than the transition between X-ray attenuation coefficients in the CT object. The goal of this work is to modify the idea of exploiting gradient-magnitude sparsity to allow for smoother transitions between regions of approximately constant values of tracer concentration.

This paper proposes an iterative algorithm for few-view SPECT reconstruction that allows for smoothed step-like variation within the object by phenomenologically modeling the SPECT object as a blurred version of a piecewise constant object. Using this model, a first-order primal-dual technique is implemented as an iterative procedure (Chambolle and Pock 2011, Sidky *et al* 2012). The purpose of this study was to characterize the performance of the algorithm under varying sampling and noise conditions, including cases where the object does not match the phenomenological model. Images reconstructed by Maximum-Likelihood Expectation Maximization (MLEM) serve as a reference. The article is organized as follows: Section 2 provides the image reconstruction theory and algorithm. Sections 3 and 4 demonstrate the algorithm with data generated using the system matrix and with data generated from a realistic Monte Carlo simulation of a SPECT system, respectively. Section 5 summarizes the results.

## 2. The algorithm

The iterative image reconstruction algorithm (IIR) is designed by defining an optimization problem which implicitly specifies the object function based on a realistic data model and a model for object sparsity. In this preliminary investigation, the specified optimization problem is solved in order to characterize its solution and the solution's appropriateness for few-view/dynamic SPECT imaging. Future work will consider algorithm efficiency by designing IIR for approximate solution of the proposed optimization problem.

### 2.1. The SPECT optimization problem

The proposed SPECT optimization problem is formulated as an unconstrained minimization of an objective function which is the sum of a data fidelity term and an image regularity penalty. The design of both terms expresses the proper SPECT noise model and a modified version of gradient-magnitude object sparsity. We first describe how standard gradient-magnitude sparsity is incorporated into a SPECT optimization problem, and then we present our modified optimization which accounts for the smoother variations expected in a SPECT object function.

*2.1.1. Unconstrained minimization for gradient-magnitude sparsity exploiting SPECT IIR.* In expressing the SPECT data fidelity term, the data are modeled as a Poisson process the mean of which is described by the following linear system of equations:

$$\mathbf{g} = \mathbf{Hf} \qquad (1)$$

where $\mathbf{H}$ is the system matrix that describes the probability that a photon emitted from a certain location in the object vector, $\mathbf{f}$, contributes to the measured data vector, $\mathbf{g}$, at a certain location. Iterative tomographic image reconstruction techniques such as MLEM and Ordered Subset Expectation Maximization (OSEM) maximize the log-likelihood of this Poisson random variable (Shepp and Vardi 1982, Hudson and Larkin 1994, Vandenberghe *et al* 2001). This is equivalent to minimizing the Kullback-Leibler (KL) data divergence ($D_{\mathrm{KL}}$) (Barrett and Myers 2004). For the present application of few-view SPECT, the data are acquired over too few views to provide a unique maximum likelihood image. In the limit of infinite photon counts and assuming that the mean model in (1) perfectly describes the imaging system, the underlying object function still cannot be determined because (1) is underdetermined.

In order to arrive at a reasonable solution, additional information or assumptions on the object function are needed. Recently, exploitation of gradient-magnitude sparsity has received much attention and has been implemented in IIR for few-view CT (Chen *et al* 2008, Sidky *et al* 2009). This idea is an example of a general strategy under much recent investigation in CS, where sampling conditions are based on some form of identified sparsity in the image. In our application the strategy calls for narrowing the solution space to only images that exactly solve our linear model in (1). Among those images, the solution with the lowest TV is sought. In practice, this solution can be obtained approximately by combining a data fidelity term with a TV penalty, where the combination coefficient in front of the TV penalty is vanishingly small. The TV-$D_{\mathrm{KL}}$ sum yields the following minimization:

$$\mathrm{minimize}_f \ \{D_{\mathrm{KL}}(\mathbf{g}, \mathbf{Hf}) + \gamma \|(|\mathbf{Df}|)\|_1 \}, \qquad (2)$$

where $\boldsymbol{D}$ is a discrete gradient operator and $\gamma$ is a weighting parameter. For sparsity-exploiting IIR, $\gamma$ is chosen so that the data fidelity term far outweighs the TV-term. The role of the TV term is simply to break the degeneracy in the objective function among all solutions of (1).

The success of TV minimization for few-view CT IIR relies on the assumption that the X-ray attenuation coefficient map is approximately piecewise constant. Directly promoting sparsity of the gradient-magnitude image may not be as beneficial for SPECT, as in some cases tracer uptake may vary smoothly within objects, and borders of objects may show a smoothed step-like dependence. For example, some regions of the heart are supplied by a single coronary artery while other regions are supplied by multiple coronary arteries (Donato *et al* 2012, Pereztol-Valdés *et al* 2005). Thus, cardiac perfusion studies may be one application for which the blurred piecewise constant model is appropriate. As another example, tumor vascularization is heterogeneous, with vascularization often varying from the tumor center to the periphery (Jain 1988). Therefore, our goal here is to find a sparsity-exploiting formulation which allows some degree of smoothness between regions with different uptake.

*2.1.2. Unconstrained minimization for sparsity exploiting IIR using a blurred piecewise constant object model.* In this work the TV minimization detailed in (2) is modified to allow for rapid but smooth variation by phenomenologically modeling objects as piecewise constant subject to a shift-invariant blurring operation.

The additional blurring operation can be incorporated into the framework developed above by minimizing the weighted sum of the TV of an intermediate piecewise constant object estimate and $D_{\text{KL}}$ between the measured data and the projection data of the blurred object estimate. The modified TV-minimization problem becomes:

$$\text{minimize}_f \ \{D_{\text{KL}}(\boldsymbol{g}, \boldsymbol{Hu}) + \gamma \||(|\boldsymbol{Df}|)\|_1 \}, \tag{3}$$

where $\boldsymbol{u}$ is the object estimate and $\boldsymbol{f}$ is an intermediate image with sparse gradient-magnitude. These are related by $\boldsymbol{u} = \boldsymbol{MGMf}$, where $\boldsymbol{M}$ is a support preserving image mask and $\boldsymbol{G}$ is a Gaussian blurring operation with standard deviation $r$. The operators $\boldsymbol{M}$ and $\boldsymbol{G}$ are symmetric so $\boldsymbol{M}^{\text{T}} = \boldsymbol{M}$ and $\boldsymbol{G}^{\text{T}} = \boldsymbol{G}$. The operator $\boldsymbol{G}$ extends data outside the physical support of the system assumed by $\boldsymbol{H}$ so the image mask $\boldsymbol{M}$ must be applied before and after

*G*. This optimization problem has two design parameters, γ, which is the weighting of the TV term, and *r*, which is the standard deviation of the Gaussian blurring kernel. The blurring parameter, *r*, represents smoothness in the underlying object, as opposed to blurring introduced by the imaging system. When $r = 0$, this formulation defaults to TV minimization problem in (2). If $\gamma = 0$, the formulation described by (3) minimizes $D_{KL}$, which is implicitly minimized in MLEM. The final image estimate is ***u***, the result of blurring and masking the intermediate piecewise constant object, ***f***. Minimizing (3) jointly enforces sparsity (by requiring a low TV of ***f***) and encourages data match (by requiring a low $D_{KL}$).

*2.2. Optimization algorithm*

Only recently have algorithms been developed that can be applied to large-scale, non-smooth convex optimization problems such as that posed by (3). Sidky *et al* (2012) adapts the Chambolle-Pock (CP) algorithm to solve the TV-$D_{KL}$ sum described by (2) (Chambolle and Pock 2011). Applying the model as described above, this prototype can be modified to solve the optimization posed by (3). Pseudo-code describing this algorithm is written below.

**Listing 1:** Pseudocode of the proposed algorithm
---
$L := \|(\textbf{HMGM}, \textbf{D})\|_2$; $\tau = \sigma = 0.9/L$; $\theta = 1$; $n = 0$
$f_0 := f'_0 := p_0 := q_0 := 0$
**Repeat**
  $p_{n+1} := 0.5(1 + p_n + \sigma \textbf{HMGM}f'_n - ((p_n + \sigma \textbf{HMGM}f'_n - 1)^2 + 4\sigma g)^{1/2}$
  $q_{n+1} := \gamma(q_n + \sigma \textbf{D}f'_n) / \max(\gamma, |q_n + \sigma \textbf{D}f'_n|)$
  $f_{n+1} := f_n - \tau \textbf{MGMH}^T p_{n+1} + \tau div(q_{n+1})$
  $f'_{n+1} := f_{n+1} + \theta(f_{n+1} - f_n)$
  $n = n+1$
**Until** *stopping criterion*
---

This algorithm is a modification of Algorithm 5 described in previous work by Sidky *et al* (2012). The convergence criterion described in that work was used here.

Simulation studies were conducted to characterize the performance of the proposed reconstruction technique over a range of angular sampling conditions, including cases in which the object does not match the phenomenological blurred piecewise constant model. The first simulation study used noiseless data generated from the system forward model to validate that the reconstruction technique closely approximates the true object when the correct blurring and system models are used, and to investigate the effects of the design parameters $r$ and $\gamma$. Another study reconstructed data generated by Monte Carlo simulation for a range of sampling and noise conditions and for varying values of algorithm parameters $r$ and $\gamma$.

**3. Inverse crime simulation study**

This study was designed to validate that the reconstruction technique approximates the true object when both the object model and system model are known exactly. The simulated object was generated from the object model and data were generated from the system forward model. Cases such as this, in which the data were produced directly from the model are referred to as the "inverse crime" (Kaipio and Somersalo 2005). This is investigated in the many-view (128 views) and few-view (9 views) cases. We also examine the effects of different blurring models on the gradient-magnitude sparsity of the intermediate object $f$. The algorithm could enable further reductions in sampling if the blurring model increases the gradient-magnitude sparsity compared to the conventional TV minimization term. In order to investigate the performance of the reconstruction with inconsistent data, Poisson noise was added to the data and the study was repeated. We refer to this as the "noisy" case.

*3.1. Methods*

*3.1.1. Phantom.* The intermediate piecewise constant object, $f_{\text{true}}$, was defined on a 128 x 128 grid of 1-mm x 1-mm pixels, representing a 6-mm diameter disk embedded in a 76-mm diameter disk. The intensity of the small disk was 2000 arbitrary units and the intensity of the large disk was 200 arbitrary units. A Gaussian blurring kernel with standard deviation, $r_{true}$ = 0.75 pixels was applied to this intermediate object to generate the ground-truth object. The intermediate object and the output of the blurring operation, $\boldsymbol{u}_{true}$ are shown in figure 1.

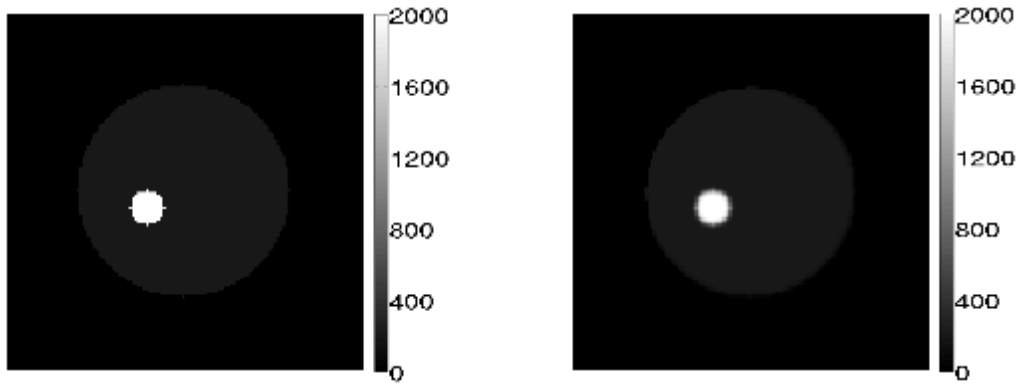

170  Figure 1. Piecewise Constant Object (left) and Phantom (right) used for simulations that generated data from the system matrix.

*3.1.2. Simulation.* Projection data of the pixelized ground-truth object was generated from the system matrix. The system matrix was estimated using Siddon's raytracing algorithm for a single-pinhole SPECT system with 3

175  mm pinhole diameter, 1.0 mm system FWHM, 35 mm pinhole-to-object distance and 63.5 mm pinhole-to-detector distance (Siddon 1985). Projection data were generated and reconstructed using 128 views, 60 views, 21 views, 15 views and 9 views, uniformly distributed around 360 degrees.

A parametric sweep was performed to investigate the effects of the two parameters on the reconstructions: the TV weighting parameter, $\gamma$, and the standard deviation of the Gaussian blurring kernel, $r$.

180  Reconstructions were performed with $\gamma$ varying from 0.0001 to 1.0 and $r$ varying from 0 to 2.0 pixels. For this case, $r_{true}$ is known to be equal to 0.75. In practice, the amount of smoothness within the underlying object is unknown and may vary across the FOV. In this study, images are reconstructed using a range of $r$ values to quantify the performance of the reconstruction technique for the expected case where the assumed $r$ differs from $r_{true}$. To reduce the necessary sampling for accurate image reconstruction, a sparse representation of an image

185  must exist. Our proposed reconstruction approach assumes that the gradient-magnitude of the intermediate image $f$ has very few meaningful coefficients. However, using an incorrect blurring model in the reconstruction may negatively affect the sparsity of the intermediate object, $f$, limiting the benefits of the algorithm. To investigate the effect of the assumed blurring model on the sparsity of the reconstructed intermediate object, $f$, images were reconstructed from 9 and 128 views using a range of $r$ values and sparsity evaluated as the number of coefficients greater than 10% of the maximum coefficient in the gradient-magnitude image of $f$.

To investigate the performance of the reconstruction technique in the presence of noise, simulations varying the number of views and parameter values were repeated with Poisson noise added to the projections generated from the system model. All simulations modeled approximately 1052000 counts, thus the peak number of counts in the 128 view projections was 298 while the peak number of counts in the 9 view projections was 3758. The noisy projection data were also reconstructed with MLEM in order to provide a reference reconstruction for comparison. As will be described in the next section, the correlation coefficient (CC) of the reconstructed image with the true object is used as a metric of accuracy throughout this work. In order to select a comparable stopping iteration for MLEM reconstruction, the CC was calculated at each MLEM iteration and the final image selected as that with the highest CC value.

*3.1.3. Metrics.* Evaluating the accuracy of the reconstructed object requires a measure of similarity or error between the reconstructed object and the true object. In SPECT imaging, including the Geant4 Application for Tomographic Emission (GATE) simulations proposed in section 4, the reconstructed activity is a scaled version of the true activity, with the scaling factor dependent on the geometric efficiency of the system (Jan *et al* 2004). Our reconstruction methods correct for the spatially varying sensitivity of the SPECT system, as will be described in section 4.1.2. However, a global scaling correction factor is not applied because absolute quantification in SPECT is challenging and may confound the characterization of the algorithm. Therefore, our accuracy metric must provide a meaningful measure of similarity in cases where the scaling factor between the reconstructed and true object is unknown. In this work, reconstruction accuracy was quantified using the correlation coefficient (CC) of the reconstructed image estimate with the true object. CC is defined as

$$CC = \frac{\sum_{k=1}^{M}(u(k)-\bar{u})(u_{true}(k)-\overline{u_{true}})}{\{\sum_{k=1}^{M}(u(k)-\bar{u})^2 \sum_{k=1}^{M}(u_{true}(k)-\overline{u_{true}})^2\}^{1/2}} \quad (4)$$

where $u_{true}$ is the true object, $M$ is the number of voxels and $u(k)$ is the reconstructed object value at voxel $k$. This metric is commonly used in image registration and is the optimum similarity measure for images that vary by a linear factor (Hill *et al* 2001). This metric allows the quantification of the accuracy of the spatial distribution of the object, without requiring absolute quantitative accuracy. CC is equal to one when the reconstructed object matches the true object. We also quantified the change in CC over the range of studied parameters ($r$ and $\gamma$), in order to quantify the sensitivity of the algorithm to parameter selections and to understand the performance of the algorithm when the assumed blurring parameter does not match the true object blur. Spatial resolution in the reconstructed images was quantified as the full-width at 10% of maximum (FW10M) of the central profile through the smaller disk. This measure was used instead of the more common full-width at half maximum (FWHM) because analysis of preliminary reconstructed images indicated that the FWHM was often accurate, even though the extent of the reconstructed object was greater than the true object. The FW10M more accurately quantified this blurring effect. The true object had a FW10M of 12 pixels. Signal-to-noise ratio (SNR) was calculated as the mean of a 3 pixel radius region in the background divided by the standard deviation of the same region.

*3.2. Results*

*3.2.1. Without Poisson noise.* We present results of the noiseless case in which the object was constructed from the object model and the projections were determined from the system matrix. The purpose of this study was to confirm that the reconstruction algorithm closely approximated the true object in the noiseless inverse crime case and to examine the effects of the design parameters r and γ as the number of views decreased. Both design parameters were varied and the number of angular samples reduced from 128 views to 9 views.

Reconstructions from 128 angular positions are shown in figure 2, with profiles of selected reconstructions shown in figure 3. Figure 4(a) presents plots of CC over the range of studied $\gamma$ and $r$ values

235    Reconstruction accuracy (CC) is high (CC > 0.980) for reconstruction using $\gamma < 1.0$, with CC varying by less than 2% for all $r$ investigated. Using $\gamma = 0.0001$ or $\gamma = 0.01$ and using $r = r_{true} = 0.75$, the object is recovered nearly exactly with CC exceeding 0.999 in each case. The FW10M value of the true object is 12 pixels, which is correctly depicted by reconstructions using $r < 1.0$ and $\gamma < 0.1$. The profiles demonstrate decreased amplitude and increased object extent when $\gamma = 0.1$ or $\gamma = 1.0$, suggesting blurring of the object. For the lower $\gamma$ cases, ring

240    artifacts are visible when $r$ is greater than 1.0.

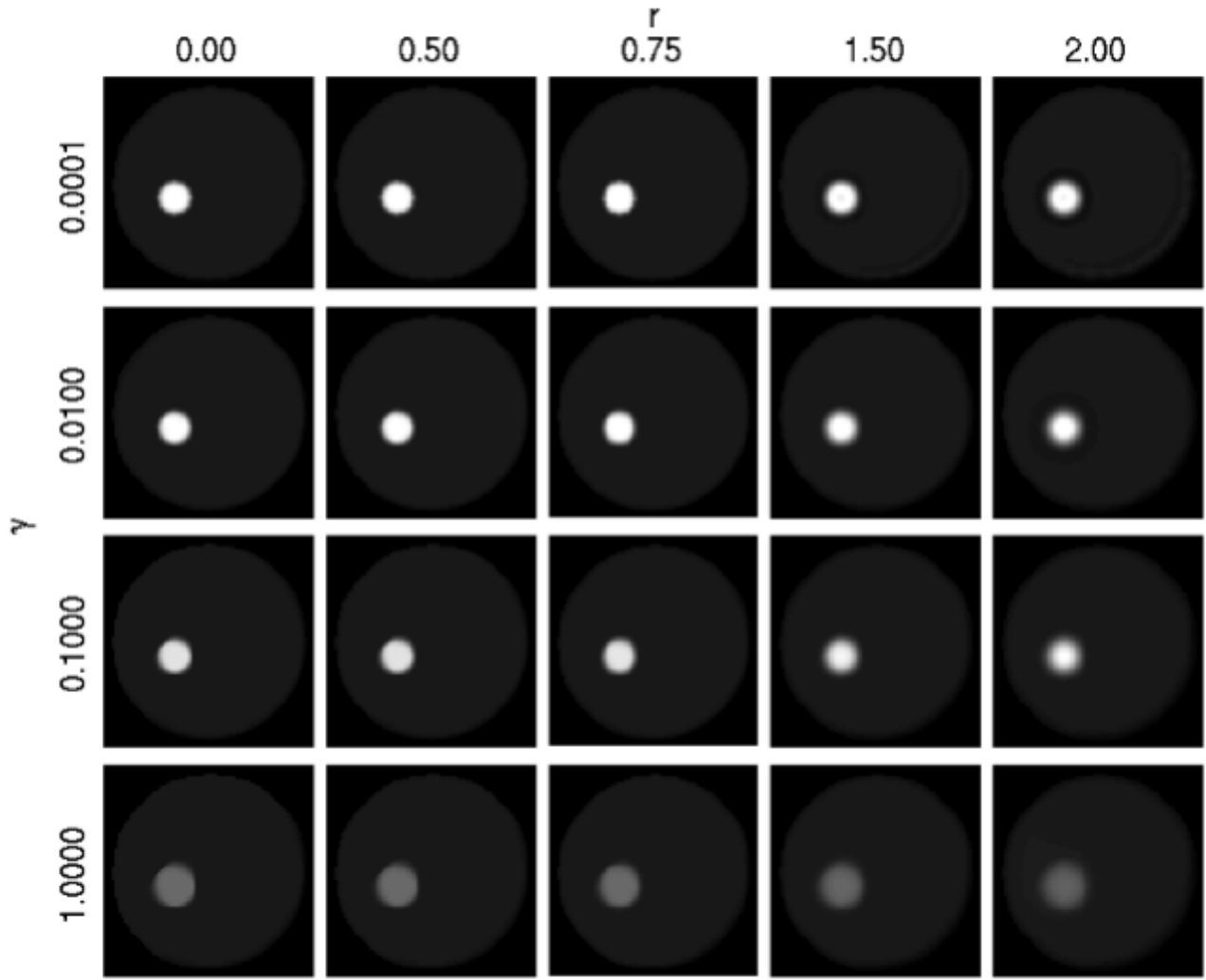

Figure 2. Images reconstructed from 128 views of noiseless inverse crime data using the proposed algorithm with varying values of r and γ.

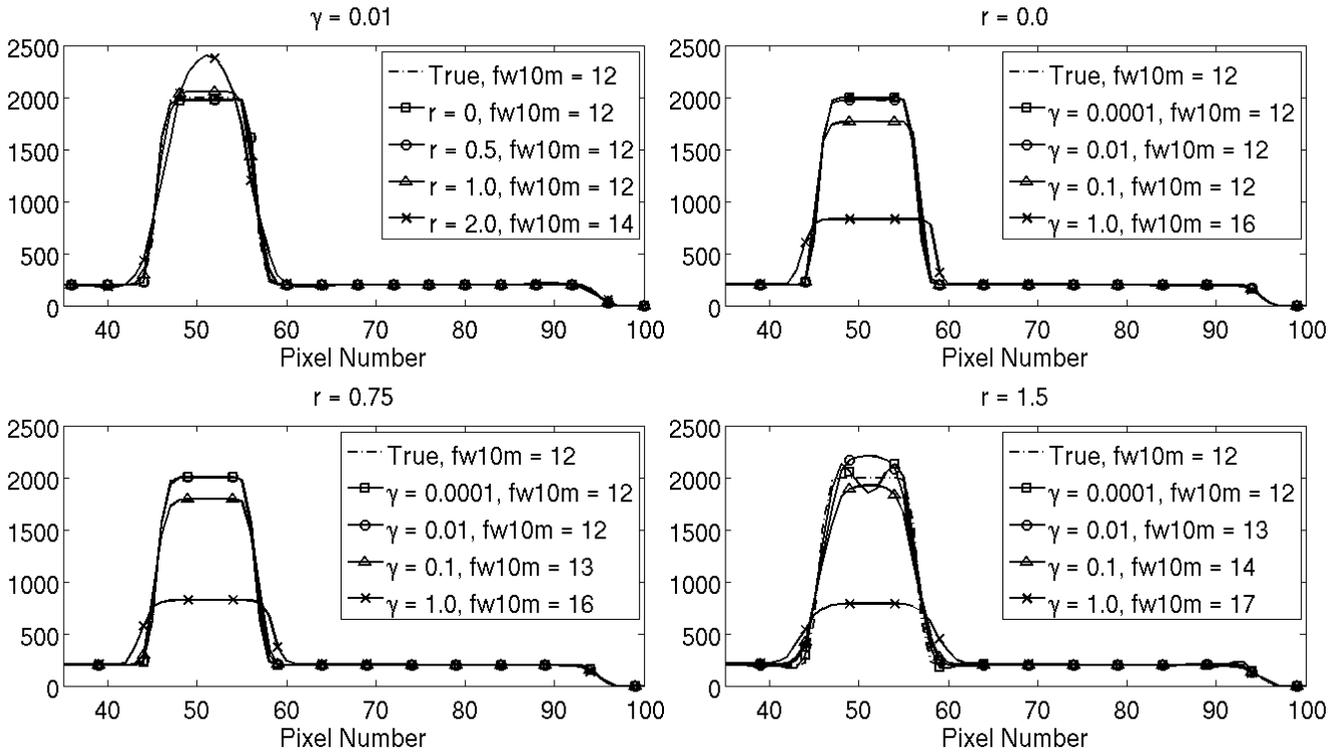

Figure 3. Central diagonal profiles through images reconstructed from noiseless inverse crime data from 128 views using the proposed algorithm with varying values of $r$ and $\gamma$.

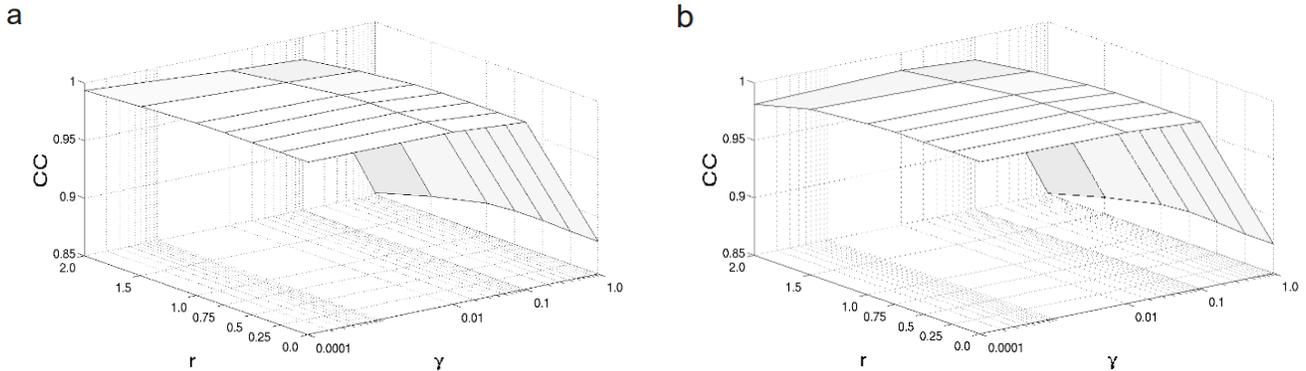

Figure 4. Plots depicting the CC over the range of studied $r$ and $\gamma$ parameters of images reconstructed from noiseless inverse crime data from 128 views (a) and 9 views (b).

The few-view case demonstrated similar trends, as shown in figures 4(b), 5 and 6. The object is nearly exactly recovered when $\gamma = 0.01$ and $r = 0.75$ is used. Using $\gamma = 0.01$, the CC varied by 1.5% (CC ranging from 0.984 - 0.999) across the range of studied $r$ values. Over the parameter set studied, CC varied between 0.878 and 0.999 depending on the value of $\gamma$ used in reconstruction, with higher $\gamma$ values resulting in lower CC. In addition to lower CC, images reconstructed with γ = 1.0 demonstrated reduced contrast and increased FW10M results, suggesting increased blurring. The FW10M value of the true object was 12, which was depicted in all reconstructions with $r \leq 1.0$ and $\gamma < 0.1$. As $r$ increased beyond 1.0, the peak value increased and the profiles demonstrate larger spread, resulting in the lowest CC for reconstructions with $r = 2.0$. Overall, in both the 128 and 9-view case, CC values demonstrated a larger range over the set of γ values compared to $r$ values, suggesting that the reconstruction technique is more sensitive to the selection of γ than $r$.

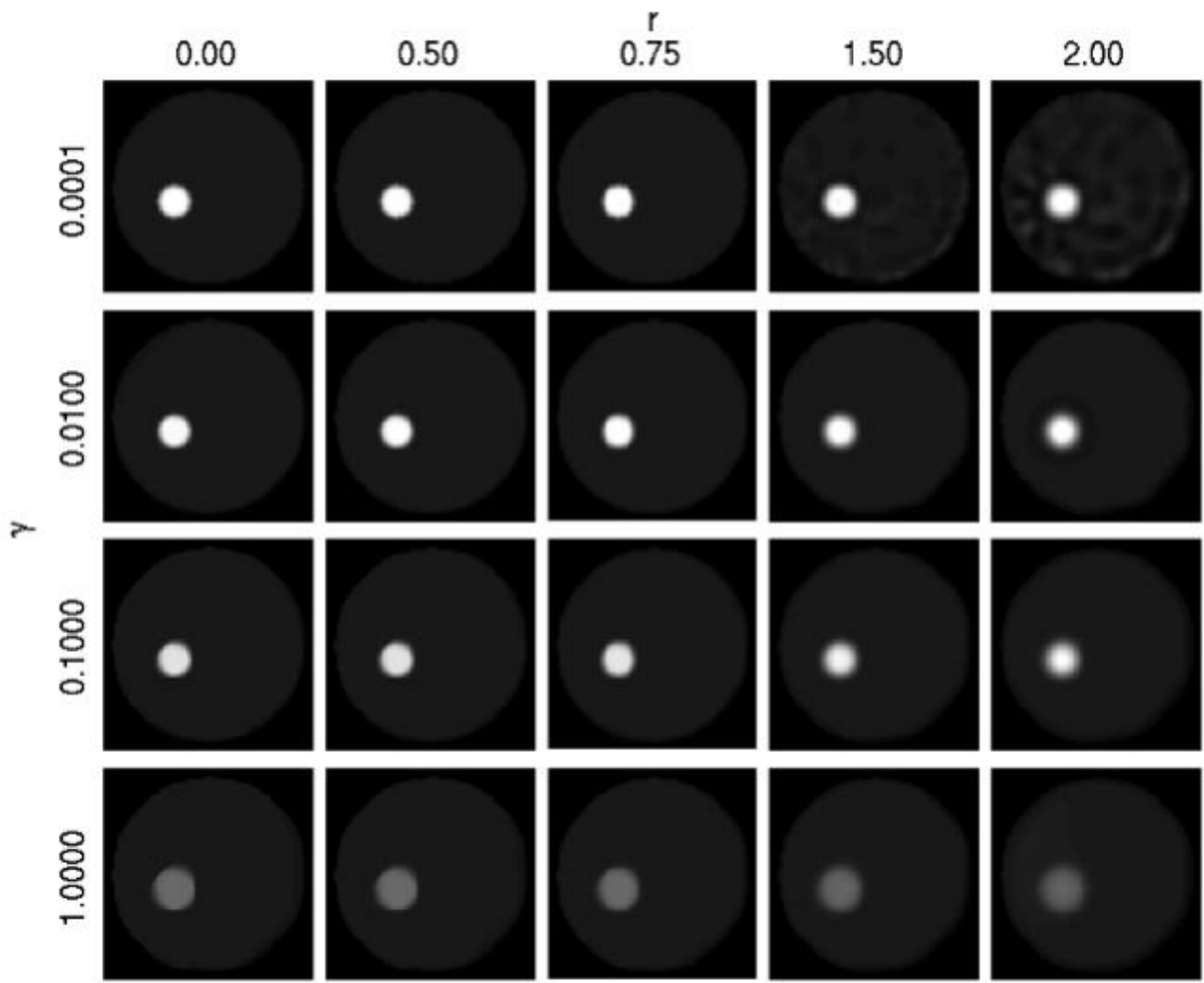

Figure 5. Images reconstructed from 9 views of noiseless inverse crime data using the proposed algorithm with varying values of r and γ.

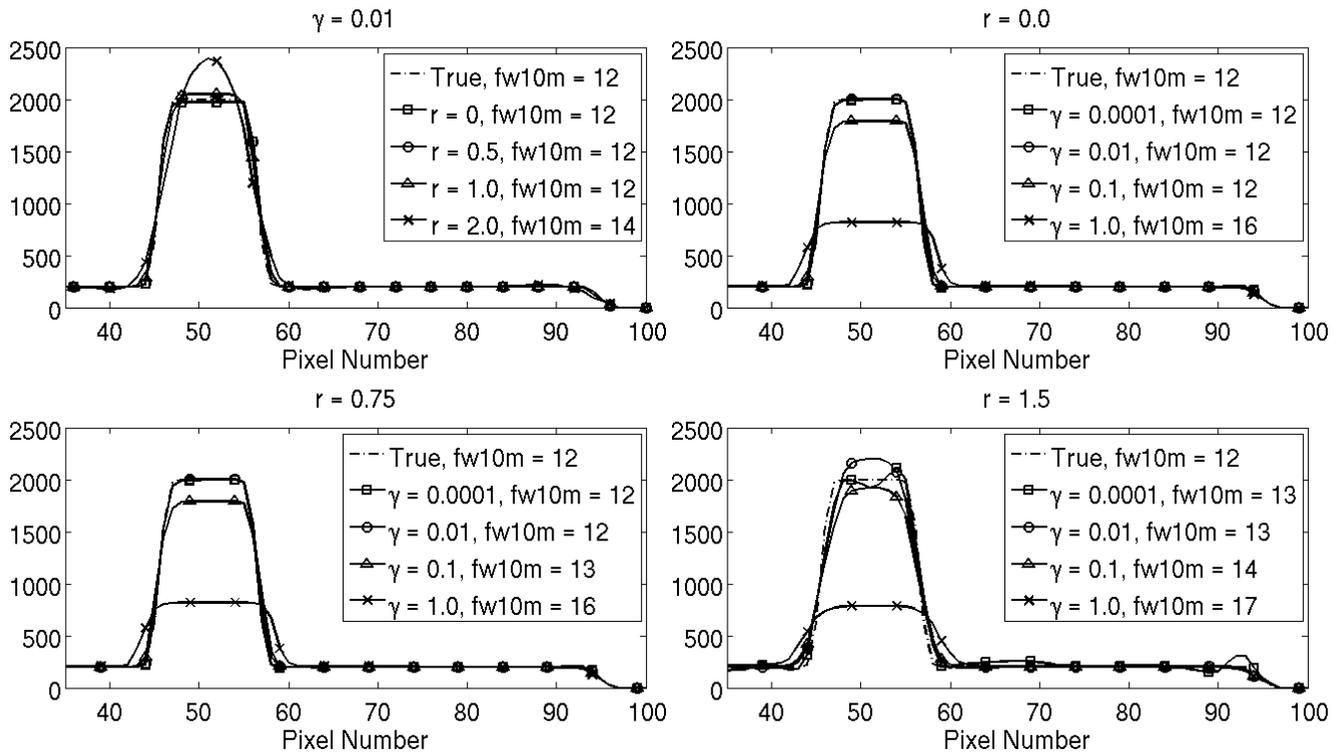

Figure 6. Central diagonal profiles through images reconstructed from noiseless inverse crime data from 9 views using the proposed algorithm with varying values of $r$ and $\gamma$.

265

*Evaluating gradient magnitude sparsity of the intermediate image.* This section evaluates the sparsity of the intermediate image $f$ reconstructed from many-view and few-view data. Figure 7 shows images of the intermediate image, $f$, reconstructed from both 128 views and 9 views using different $r$ and $\gamma = 0.0001$. Each image is captioned by its sparsity value (number of meaningful coefficients).

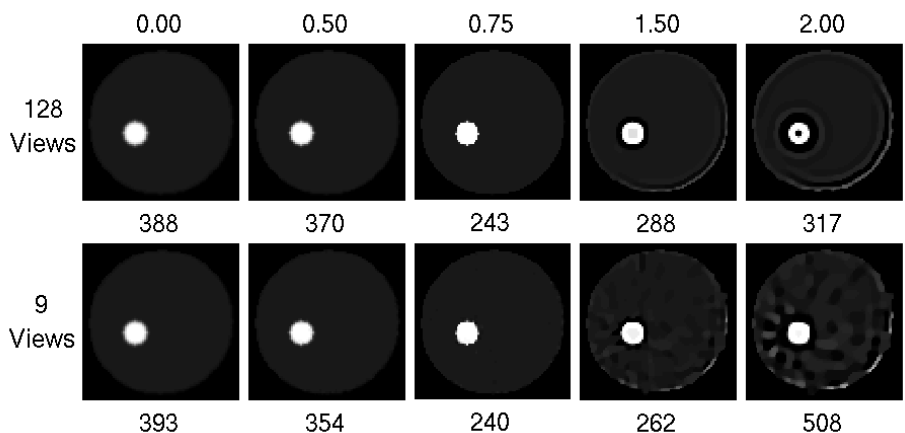

270	Figure 7. Intermediate images $f$ and the number of meaningful sparsity coefficients reconstructed from 128 and 9 noiseless inverse crime data using $\gamma = 0.0001$.

In both the many-view and few-view case, the image reconstructed with the true blurring model ($r =$ 0.75) was the most sparse and as the $r$ assumed by the algorithm diverged from $r_{true}$ the images became less sparse. This indicates that using the correct blurring model may allow the greatest sampling reductions. Additionally, underestimating $r$ leads to a gradual increase in the number of meaningful coefficients. In the few-view case, over-estimating $r$ leads to a rapid increase in the number of meaningful coefficients, reflected by the fact that new structure enters the image. These artifacts survive the blurring with $G$, leading to artifacts in the presented image $u$.

*3.2.2. With Poisson noise added.* We next considered data generated by the system matrix with the addition of Poisson noise. The purpose of this study was to examine the effects of noise on the reconstructions, using data from an otherwise inverse crime case, in which the object model and system model are known exactly. Figure 8 shows images reconstructed from 128 views over the range of $r$ and $\gamma$ parameters, with profiles plotted in figure 9. Figure 10(a) shows the plot of the CC metric over the range of studied parameters. As in the noiseless case, the CC varied by less than 1.5% across the studied $r$ values for $\gamma > 0.0001$. Unlike the noiseless case, when $\gamma = 0.0001$, the CC increased from 0.867 to 0.988 as $r$ increased from 0.0 to 2.0, as the increased blurring provided additional regularity and noise reduction. Noise is also reduced as $\gamma$ is increased, due to the increased weighting of the TV term. The highest CC value of 0.999 occurred when $r = 0.75$, the true value of $r$, and $\gamma = 0.01$. As in the noiseless case, contrast and spatial resolution decreased with increasing $\gamma$. The FW10M ranged from 12-14 for $\gamma < 1.0$, compared to a true value of 12.

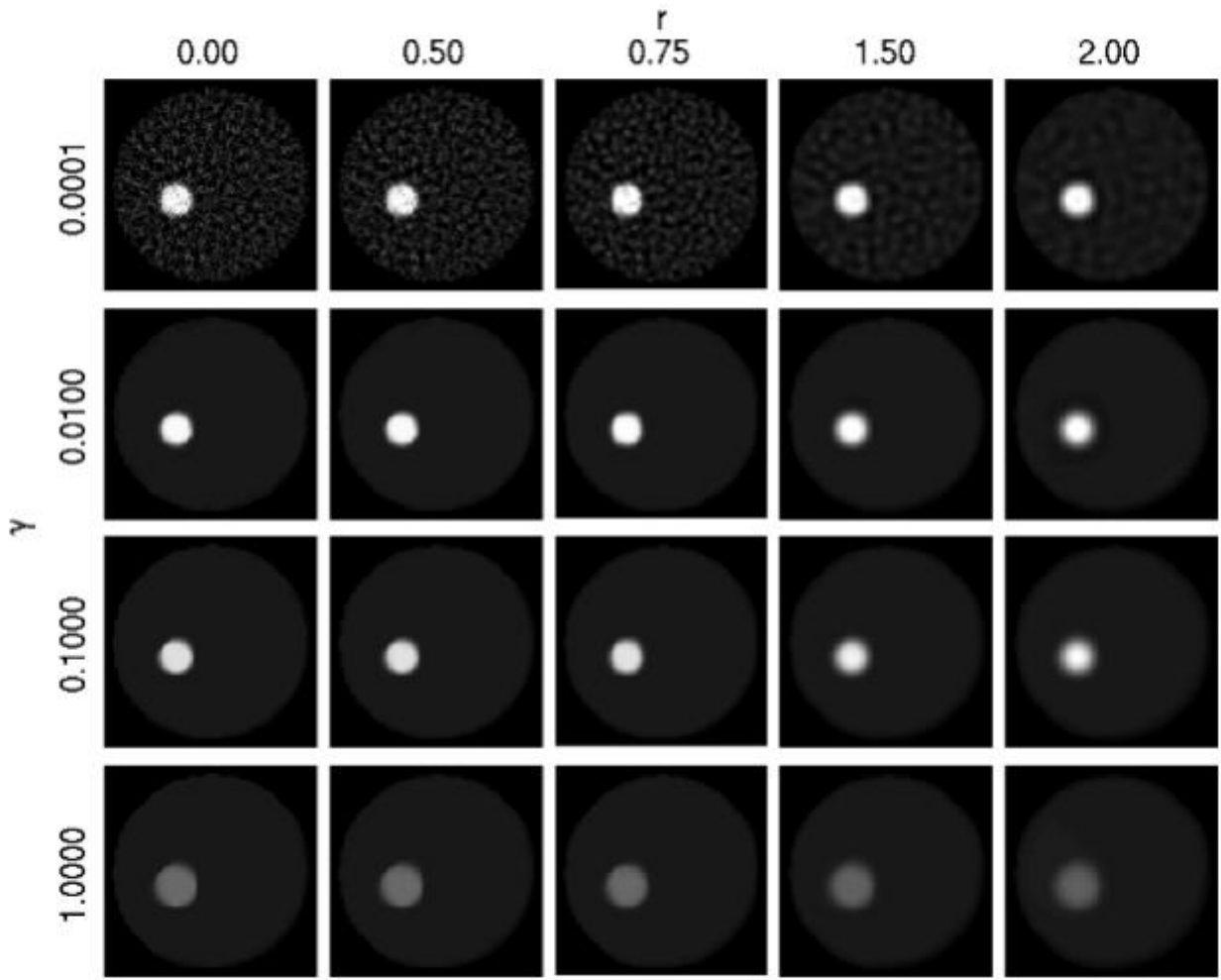

Figure 8. Images reconstructed from 128 views of noisy data using the proposed algorithm with varying values of r and γ. For these images, the projection data were generated by the system matrix.

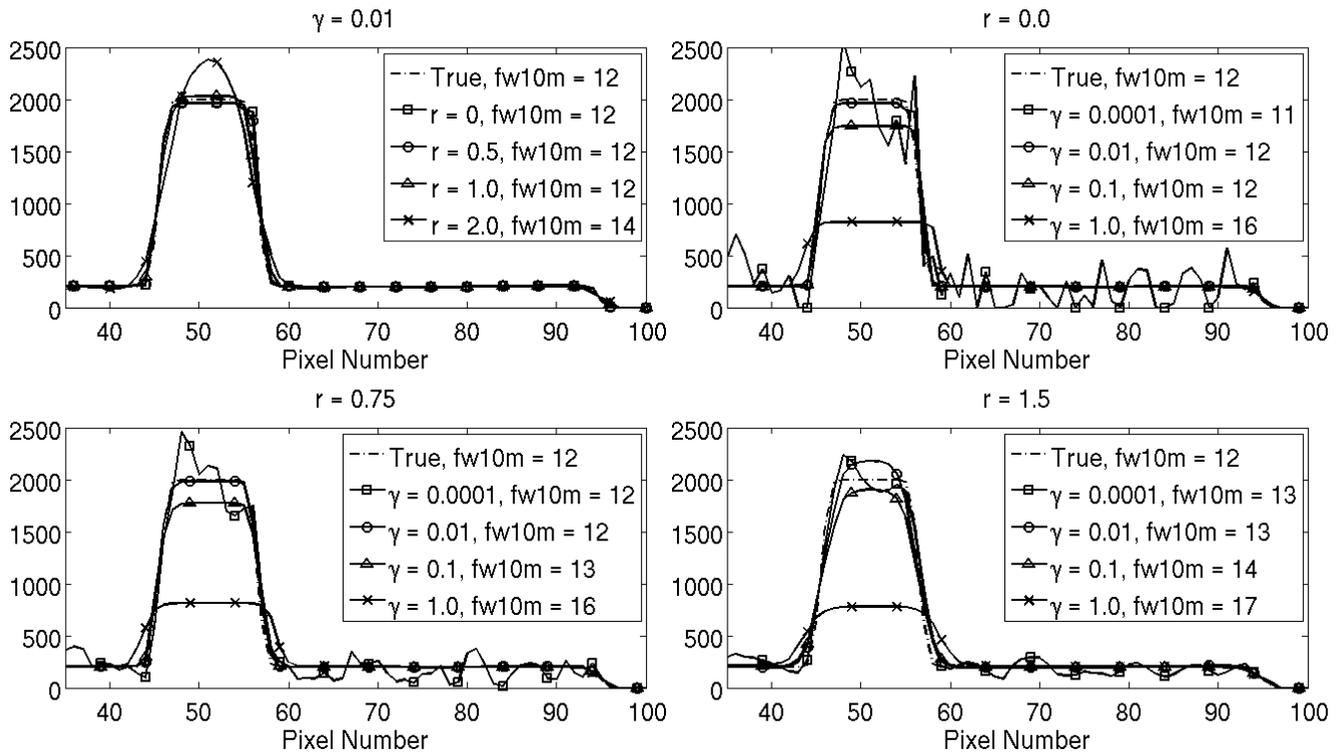

Figure 9. Central diagonal profiles through images reconstructed from noisy inverse crime data from 128 views using the proposed algorithm with varying values of $r$ and $\gamma$. For these images, the projection data were generated by the system matrix.

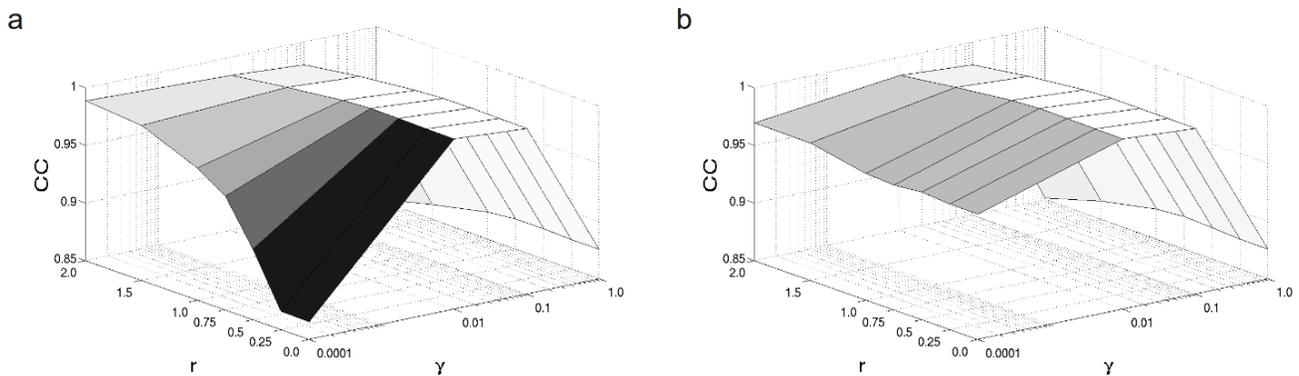

Figure 10. Plots depicting the CC over the range of studied $r$ and $\gamma$ parameters of images reconstructed from noisy data from 128 views (a) and 9 views (b). For these images, the projection data were generated by the system matrix.

Figures 10(b), 11 and 12 display the images, profiles and plots for noisy images reconstructed from nine views. Similar trends were observed as in the reconstructions from 128 views. Images reconstructed with low $\gamma$ values ($\gamma = 0.0001$) demonstrated increased noise and streaking artifacts, which were reduced with increasing $r$.

For $\gamma = 0.01$, the highest CC occurred when the assumed blurring model match the true object ($r = 0.75$), with CC varying by less than 1.4% across the range of $r$ values. The highest CC value of 0.997 was obtained with $\gamma = 0.01$ and $r = 0.75$. Similar FW10M results were obtained using 128 views, with the exception of increased FW10M (14-15) when $\gamma = 0.0001$.

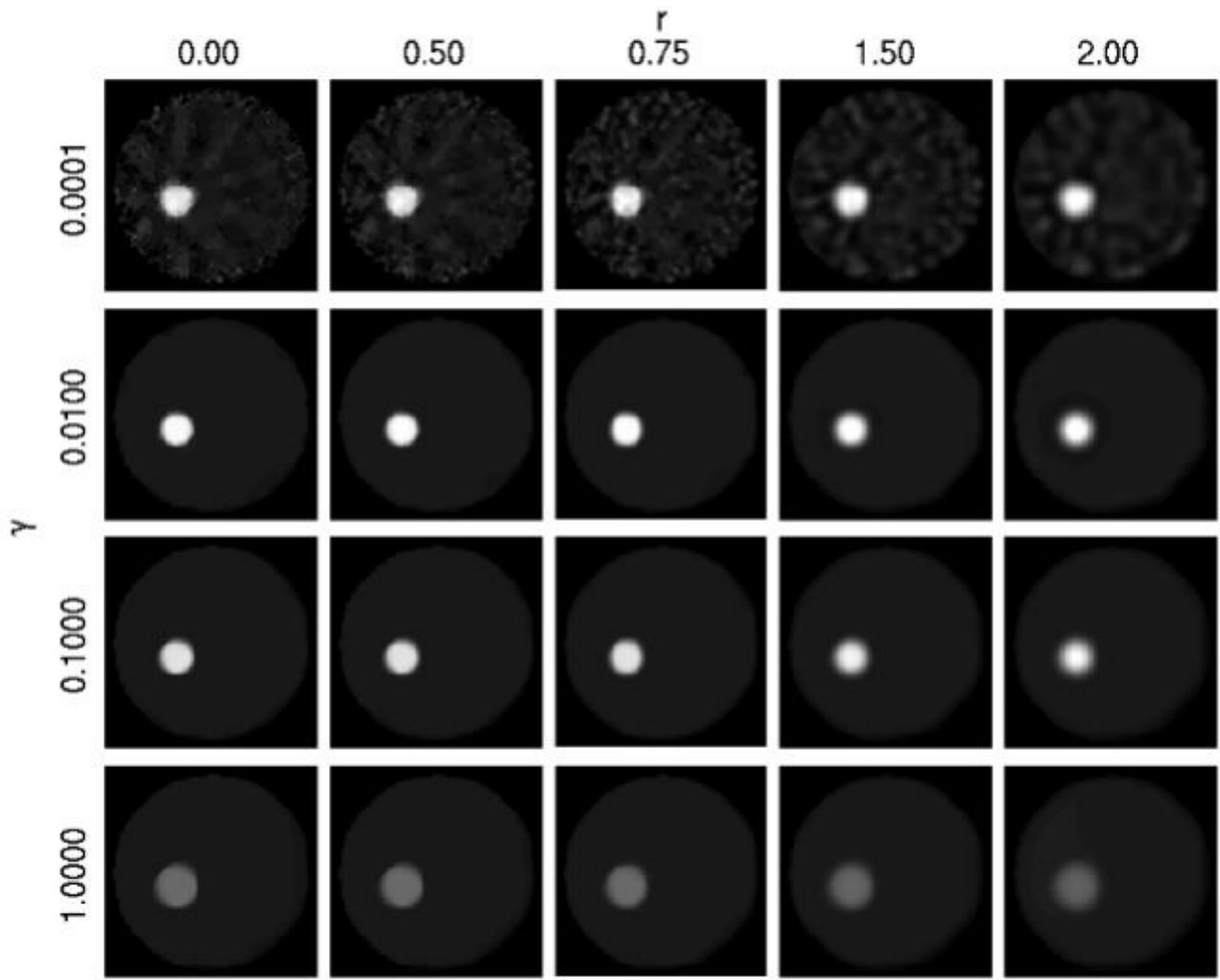

Figure 11. Images reconstructed from 9 views of noisy data using the proposed algorithm with varying values of $r$ and $\gamma$. For these images, the projection data were generated by the system matrix.

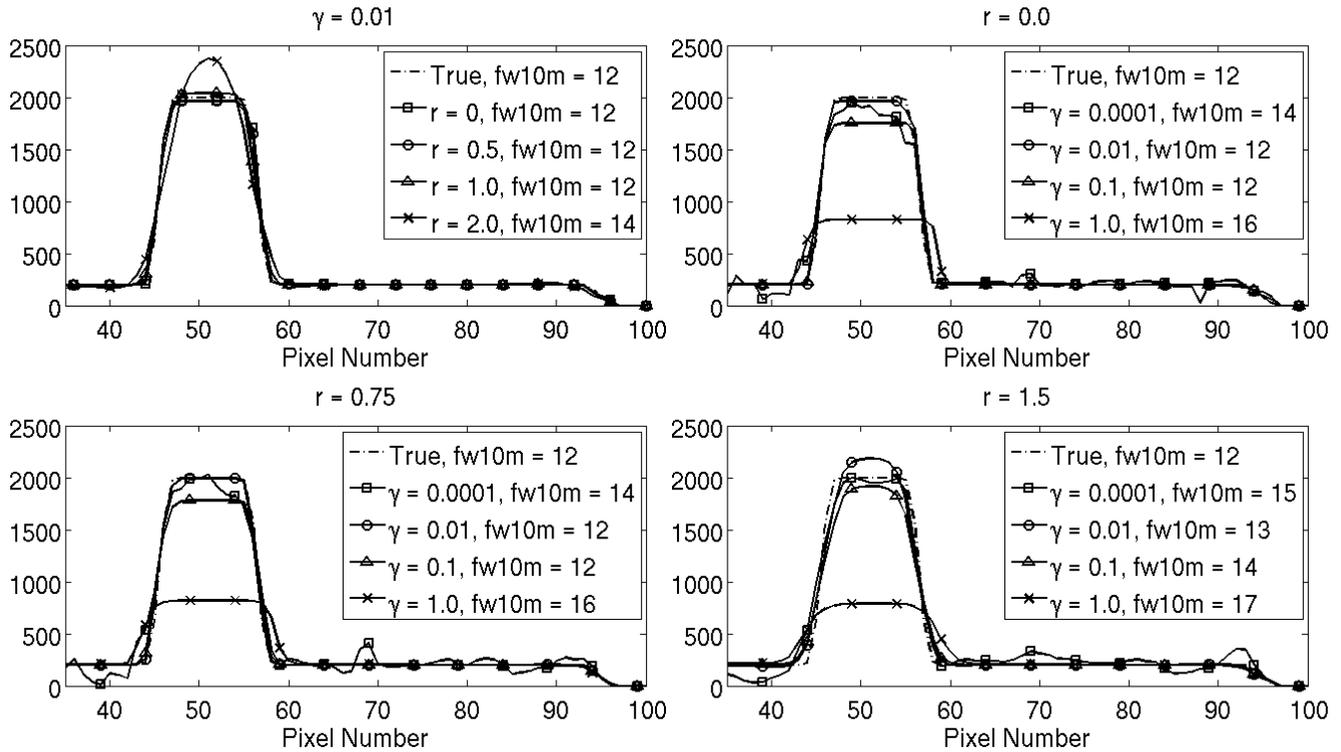

Figure 12. Central diagonal profiles through images reconstructed from noisy inverse crime data from 9 views using the proposed algorithm with varying values of $r$ and $\gamma$. For these images, the projection data were generated by the system matrix.

Overall, as in the noiseless case, CC showed little variation with $r$ but greater variation with $\gamma$, and both the 9- and 128-view reconstructions suggest that $\gamma = 0.01$ provides the highest CC. The lowest CC value in both the high- and few-view cases occurred with large values of $r$ and $\gamma$ ($r = 2.0$ and $\gamma = 1.0$).

Figure 13 compares images reconstructed with the proposed reconstruction technique ($\gamma = 0.01$ and $r = 0.75$) and MLEM from data acquired with a varying number of angular views. Table 1 shows CC, SNR and FW10M values for each reconstruction technique and number of views. Images reconstructed using the proposed algorithm had CC values that were 2-4% higher for each case compared to MLEM. The greatest difference is noted for the 9 view case where the image reconstructed using the proposed algorithm yielded a CC of 0.994 while the MLEM image had a CC of 0.954. Streak artifacts were present in the MLEM reconstructions, and were primarily absent in images reconstructed with the proposed algorithm. The noise level in MLEM reconstructions is higher, leading to lower SNR. Both algorithms provide similar FW10M values compared to the true value of 12.

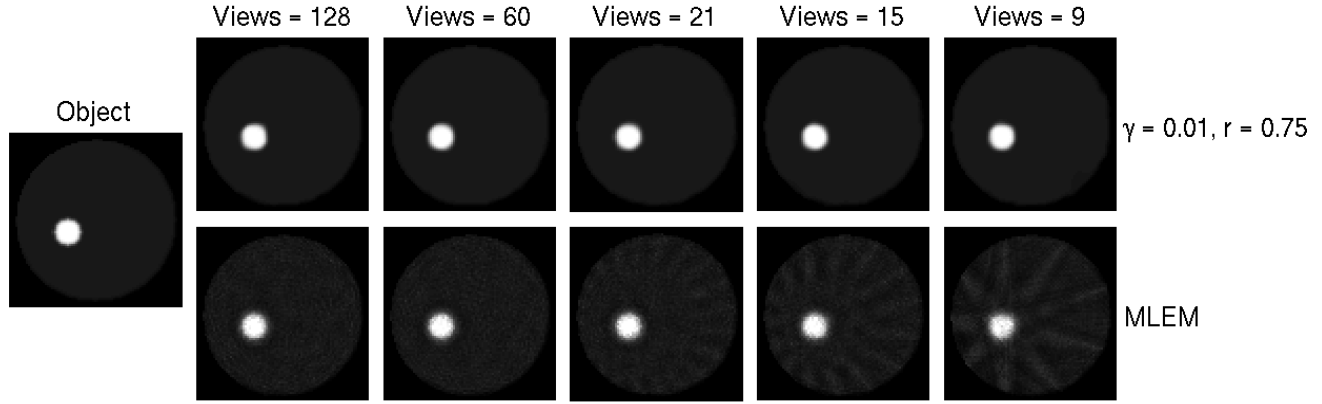

Figure 13. Images reconstructed from noisy projections using the proposed algorithm and MLEM for varying sampling cases. For these images, the projection data were generated by the system matrix.

Table 1. Comparison of image quality metrics from images reconstructed from noisy projections generated by the system matrix.

|  |  | 128 views | 60 views | 21 views | 15 views | 9 views |
|---|---|---|---|---|---|---|
| $\gamma = 0.010$ $r = 0.75$ | CC | 0.999 | 0.998 | 0.998 | 0.998 | 0.999 |
|  | SNR | 314.7 | 736.5 | 154.14 | 615.16 | 61.17 |
|  | FW10M | 12 | 12 | 12 | 12 | 12 |
| MLEM | CC | 0.986 | 0.987 | 0.984 | 0.981 | 0.973 |
|  | SNR | 6.5 | 6.13 | 5.38 | 6.53 | 6.31 |
|  | FW10M | 13 | 13 | 14 | 12 | 15 |

### 4. Monte Carlo simulation study

The purpose of this study was to characterize the performance of the reconstruction technique for the more realistic case where the object does not necessarily match the model assumed in reconstruction, and the modeled system matrix is an approximation to the system that generated the data. In addition, these simulations include realistic effects such as scatter, spatially-varying pinhole sensitivity and blurring from the pinhole aperture.

*4.1. Methods*

345 *4.1.1. Phantom.* The object was defined on a 512 x 512 pixel grid of 0.25 x 0.25 mm pixels. The object consisted of a 28 mm-radius disk of background activity containing five contrast elements of varying size, shape, and intensity, as detailed in Table 2 and displayed in figure 14. Two, two-dimensional Gaussian distributions with peak intensities 638 Bq and 319 Bq and standard deviations 4 mm and 8 mm truncated to have radius 4.4 mm were embedded in the larger disk. Also included in the phantom

350 were a disk representing a cold region with radius 4.4 mm and one disk with radius 2.2 mm having constant intensity, as detailed in Table 2. None of the elements in the phantom were generated by the smoothed piecewise constant model assumed by the reconstruction algorithm, thus representing a challenging reconstruction task.

**Table 2.** GATE Phantom Specifications.

| Element | Radius (mm) | Position (mm) | Intensity (relative) |
|---|---|---|---|
| A | 28 | (0,0) | Constant; 64 Bq/pixel |
| B | 4.4 | (-13,6) | Activity = 6.8 MBq; Peak = 638 Bq/pixel; Std Dev = 4 mm |
| C | 4.4 | (6,-13) | Activity = 0.49Mbq; Peak = 319 Bq/pixel; Std Dev = 8 mm |
| D | 4.4 | (0,15) | 0 |
| E | 2.2 | (18,4) | Constant; 640 Bq/pixel |

355

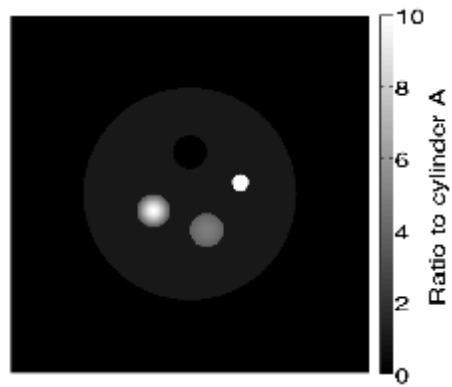

Figure 14. Voxelized phantom used in the GATE studies. The phantom contains contrast elements of varying shape and size as described in Table 2.

*4.1.2. Simulations.* Projections of the pixelized object were generated using GATE Monte Carlo simulation to model the stochastic emission of photons from a voxelized phantom and their stochastic transmission through the collimator and camera. A three-camera system was simulated. Each collimator was simulated as a 20 mm thick tungsten plate having a 3 mm diameter pinhole with 1.5 mm channel length. A 128 mm x 1 mm NaI crystal was simulated and detected photons binned into 1 mm x 1 mm pixels. Compton scatter, Rayleigh scatter and photoelectric absorption were included as possible interactions for 140 keV photons. Photons detected outside the 129.5 – 150.5 keV range were rejected as scatter. Electronic noise was not modeled. The system is described in figure 15 and table 3.

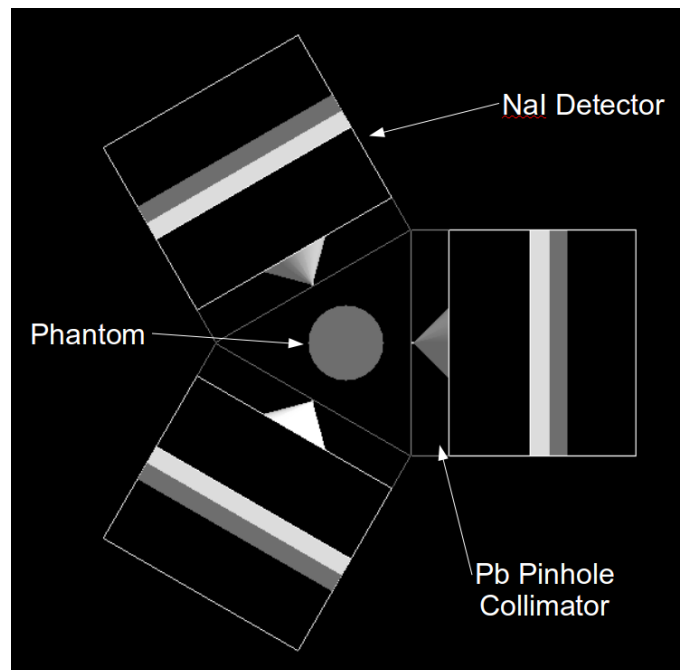

Figure 15. Diagram of Simulated SPECT system.

Table 3. Specifications of the simulated SPECT system.

| | |
|---|---|
| Camera Size | 128 mm x 120 mm x 1 mm |
| Pinhole diameter | 3 mm |
| Pinhole-to-object distance | 35 mm |
| Pinhole-to-detector distance | 63.5 mm |

370    The sensitivity of pinhole collimators depends on the angle of the ray incident on the pinhole. In order to correct for the spatially-varying pinhole sensitivity during reconstruction, a sensitivity map was generated by simulating a flood source on the collimator surface (Vanhove *et al* 2008). The resulting projection represents the spatially-varying sensitivity of the pinhole and was incorporated into the reconstruction algorithm. The sensitivity map was multiplied during each forward projection prior to the summing of data from each ray. Data

375    were multiplied by the sensitivity map prior to backprojection.

Two distinct cases were simulated. In the first case, the total simulated scan time was held constant as the number of views decreased in order to examine the effects of angular undersampling independent of changes in noise. Scans comprising 60, 21, 15, and 9 views distributed over 360 degrees acquired during a 200 second scan were simulated. These data had approximately the same number of total counts in each simulation (~65000

380    counts). The noise level in SPECT imaging is dependent on the number of detected counts, so the reconstructed images should have similar noise statistics regardless of the number of view angles. The second simulated case held the acquisition time of each view constant across all angular sampling cases. By doing so, the scans that used fewer views had improved temporal sampling, but fewer counts. As the number of views decreased, so did the absolute intensities of the reconstructed images. Images were acquired over 10 seconds for each position of

385    the three-camera gantry, thereby varying the total scan time from 200 seconds for 60 views to 30 seconds for 9 views. In this case, the simulated scan with the fewest views (9) had the fewest counts (~10000 counts) and, consequently, the highest noise level. This represents a more realistic approach for providing dynamic scans with high temporal sampling.

The simulated phantom cannot be described using a constant $r$ across the spatial domain. Each disk has

390    a definite edge and distinct profile. To investigate the effects of varying $r$ in the case where its optimal value is unknown, data were reconstructed using the TV case ($r = 0.0$) and varying $r$ from 0.25 to 2.0 pixels. The TV weighting parameter was varied from 0.0001 to 1.0. The resulting images were evaluated on the basis of reconstruction accuracy with the CC metric as described in section 3.1.3. Spatial resolution was quantified by considering FW10M of a profile through the center of disk E. A 3 pixel-radius region in the background of disk A was used to calculate SNR. In each case, MLEM reconstructions are also presented as a reference, with the MLEM stopping iteration selected as the iteration with the highest CC value.

*4.2. Results*

In this section we present the results of the Monte Carlo simulations performed over a range of angular sampling schemes for two different cases: constant total scan time and constant scan time per view.

*4.2.1. Constant total scan time.* Data reconstructed using the proposed algorithm from 60 views and a variety of $r$ and $\gamma$ values are presented in figure 15, with profiles presented in figure 17, and plots of CC in figure 18.

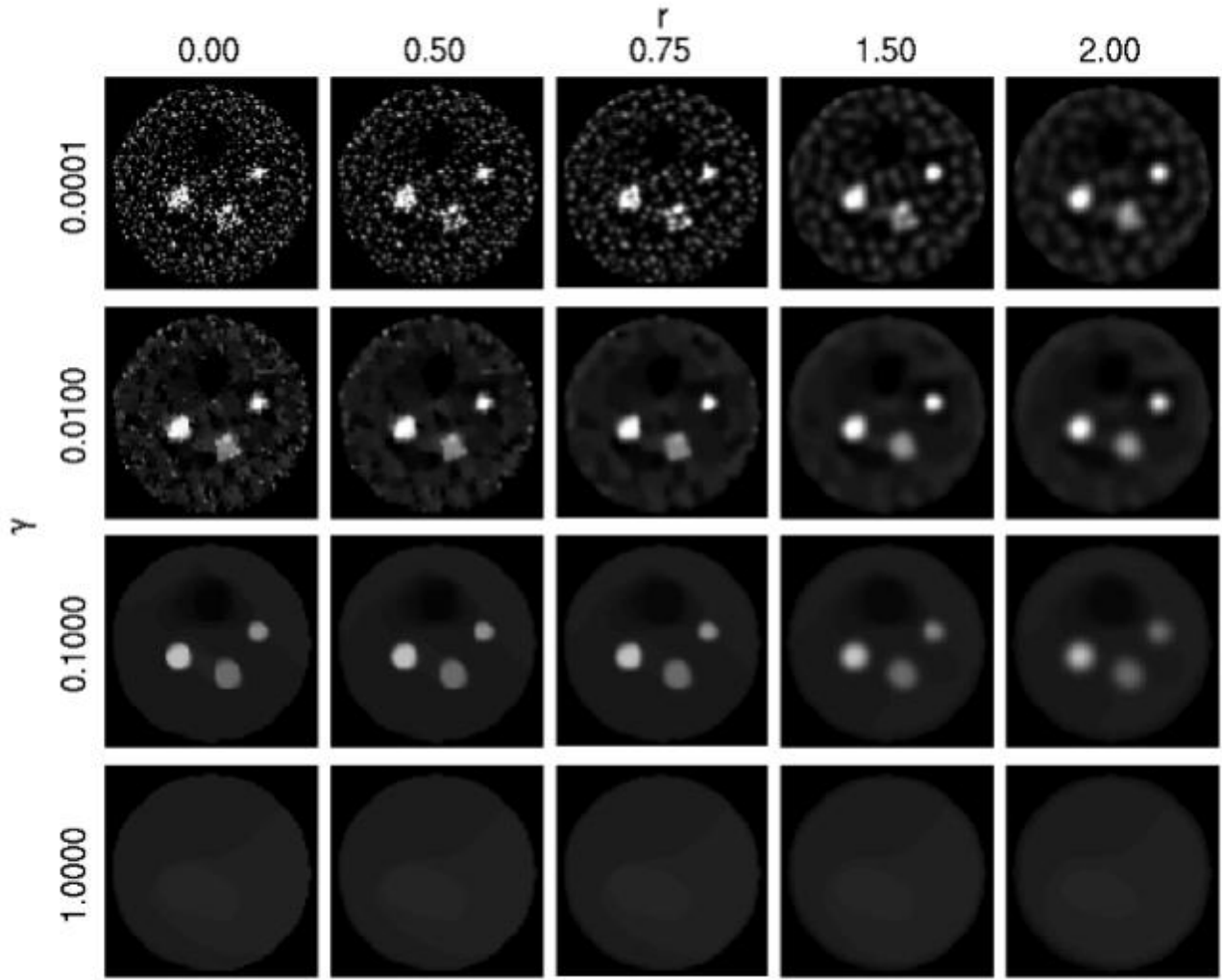

Figure 16. Images reconstructed from 60 views of GATE data simulated for 200 seconds using the proposed algorithm with varying values of r and γ.

When the TV weighting parameter was small ($\gamma = 0.0001$), the resulting image contained high frequency noise; when the TV weighting parameter was large ($\gamma = 1.0$), the object was blurred and contrast reduced. The remainder of the results will focus on $\gamma = 0.1$ and $\gamma = 0.01$. With $\gamma = 0.1$, the CC of the images with the true object varied by 3% across the studied *r* values, with CC equal to 0.942 at $r = 0.75$ and CC = 0.910 at $r = 2.0$. When $\gamma = 0.1$, the reconstructed profiles do not reach the true peak level for any values of *r*, as demonstrated in figure 17. Using $\gamma = 0.01$, the profiles reach a higher peak but the CC of these images varies by 11% from 0.946 when $r = 1.5$ to 0.836 at $r = 0.0$. As seen in figure 17, the MLEM reconstructions also did not reach the peak values of the true object profiles, suggesting that this error may be caused by system blurring rather than the reconstruction algorithm. Using $\gamma = 0.01$, the $r$ value that yielded the optimal image (in terms of CC) was 1.5, compared to an optimal $r$ value of 0.75 when $\gamma = 0.1$. The FW10M for the image reconstructed using $\gamma = 0.01$ and $r = 1.5$ was 7 pixels, compared to a FW10M of 8 pixels resulting from MLEM reconstruction, and a true value of 4. The FW10M for the images reconstructed with $\gamma = 0.1$ and $r = 0.75$ was 8 pixels.

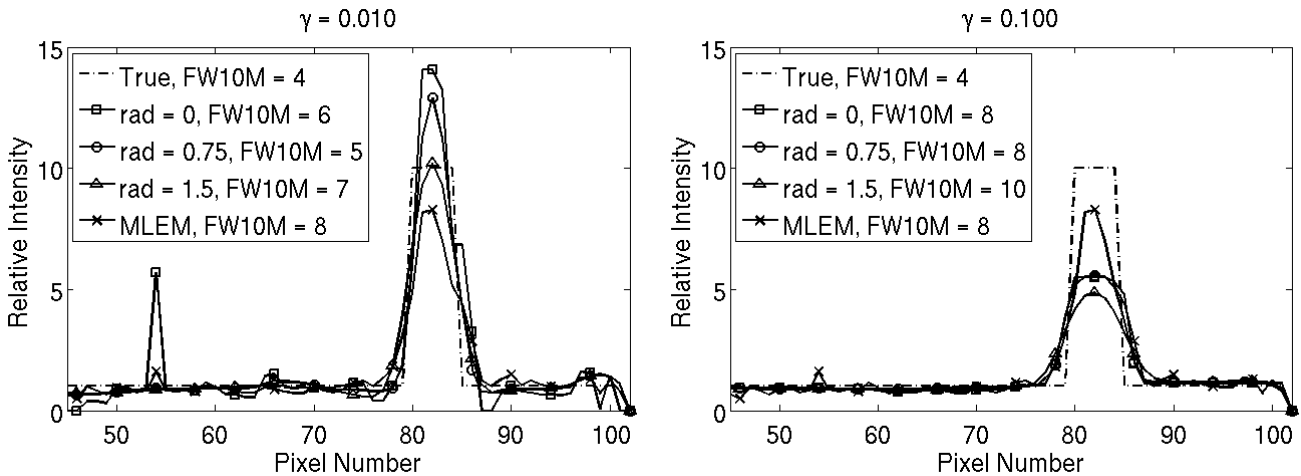

Figure 17. Central vertical profiles through images reconstructed from 60 views of GATE data simulated for 200 seconds using the proposed algorithm with varying values of $r$ and $\gamma$.

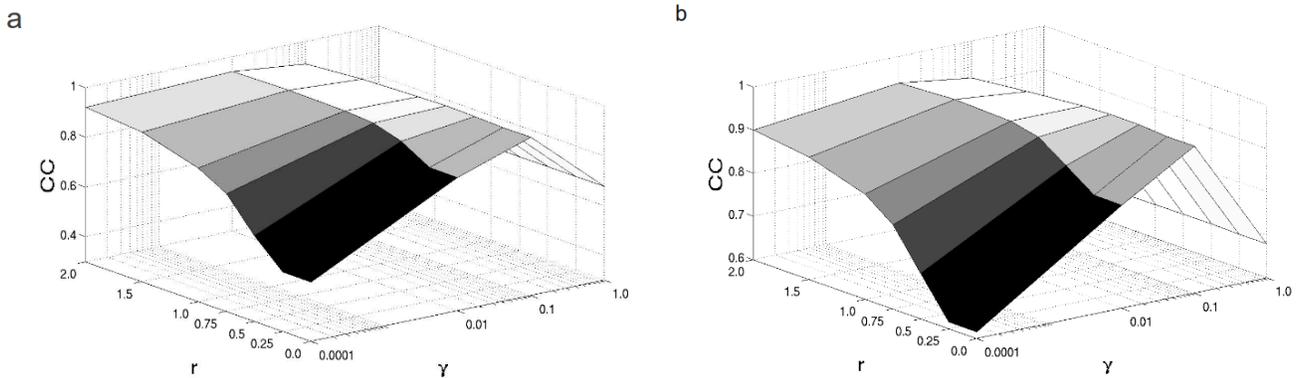

Figure 18. Plots depicting the CC over the range of studied $r$ and $\gamma$ parameters of images reconstructed from GATE data simulated for 200 seconds, using 60 views (a) and 9 views (b).

The images reconstructed from 9 views demonstrated behavior similar to images reconstructed using 60 views. Images are shown in figure 19. Images reconstructed using $\gamma = 0.01$ contained more noise than images reconstructed using $\gamma = 0.1$. The CC of images reconstructed with $\gamma = 0.01$ vary by 11.3% from 0.946 ($r = 1.5$) to 0.839 ($r = 0.0$), depending on the value of $r$. Images using $\gamma = 0.1$ had a lesser dependence on $r$, varying by 3.1% from a 0.945 peak at $r = 0.75$ to 0.915 at $r = 2.0$. Images using $\gamma = 0.01$ and $r = 1.5$ have a FW10M value of 8, compared to the true FW10M value of 4 and a FW10M of 8 resulting from MLEM reconstruction. The FW10M of images reconstructed using $\gamma = 0.1$ and $r = 0.75$ was 9 pixels. Profiles are shown in figure 20.

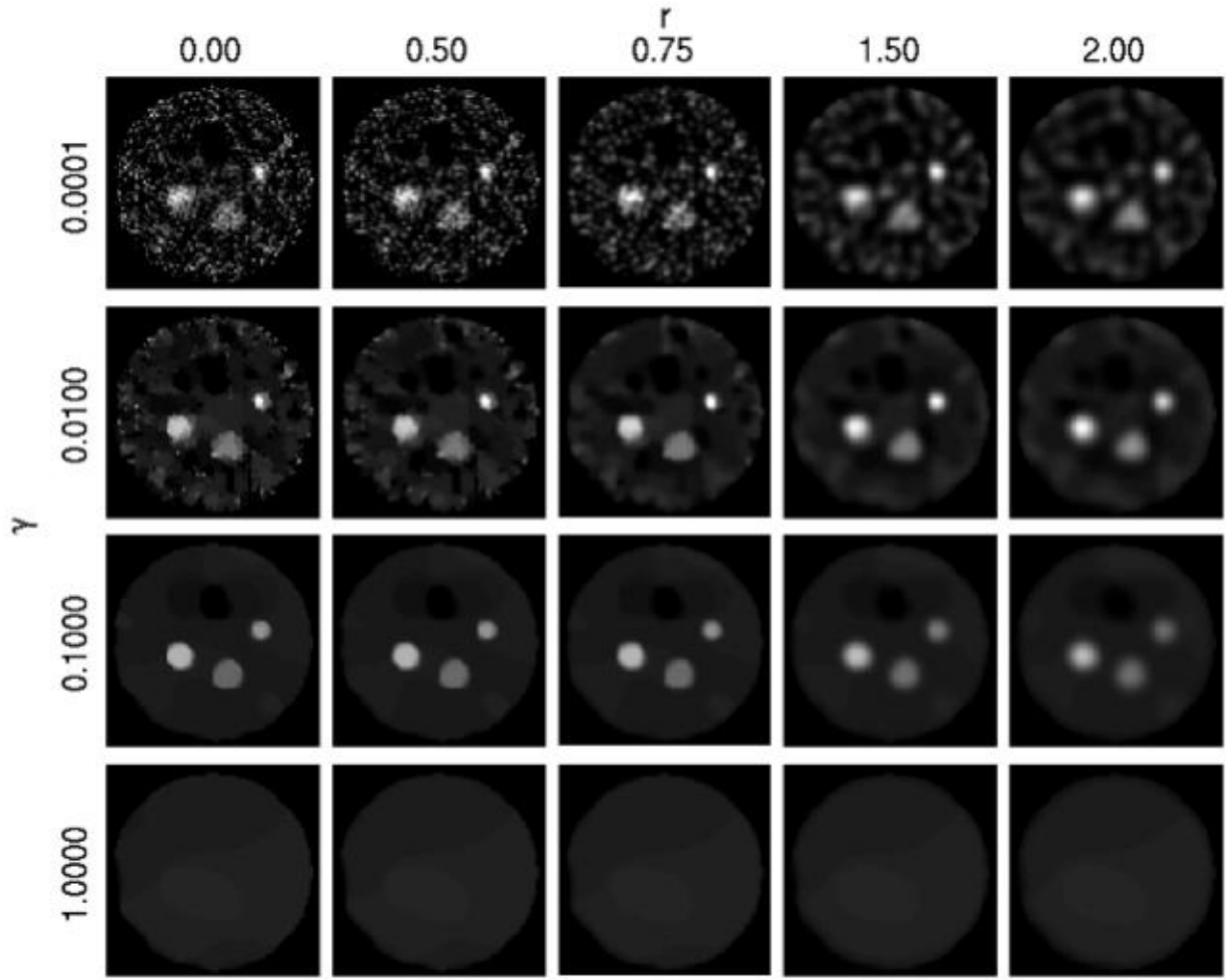

435   Figure 19. Images reconstructed from 9 views of GATE data simulated for 200 seconds using the proposed algorithm with varying values of r and $\gamma$.

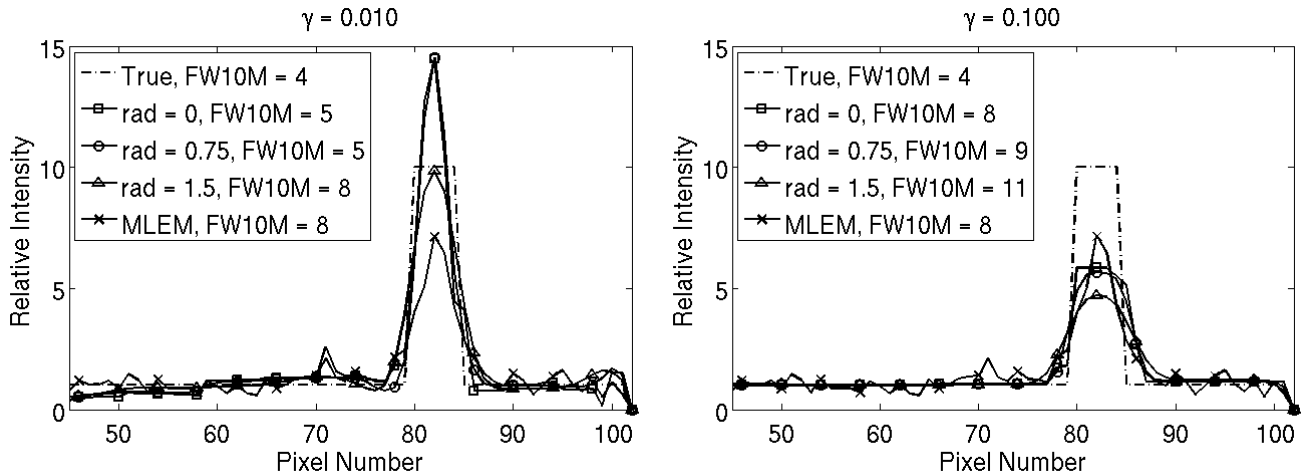

Figure 20. Central vertical profiles through images reconstructed from 9 views GATE data simulated for 200 seconds using the proposed algorithm with varying values of *r* and γ.

Figure 21 compares images reconstructed with the proposed algorithm (γ = 0.01, *r* = 1.5 and γ = 0.1, *r* = 0.75) and MLEM from data acquired with a varying number of angular views. For images reconstructed using the proposed reconstruction technique with γ = 0.01 and *r* = 1.50, the CC of the images varied by less than 1% from 0.946 to 0.942 as the number of view decreases from 60 to 9. The CC varied similarly for images reconstructed using γ = 0.1 and *r* = 0.75. For comparison, the CC of images reconstructed by MLEM decreased 6.5% from 0.913 to 0.854 as the number of views degrease from 60 to 9. For this object, the proposed reconstruction algorithm using both γ = 0.01 and γ = 0.1 provided higher CC and lower SNR compared to MLEM for all angular cases, while providing similar FW10M values. Images reconstructed using the proposed algorithm contained low-frequency patchy artifacts in the background due to noise, while reducing the streak artifacts present in MLEM reconstructions from few-views.

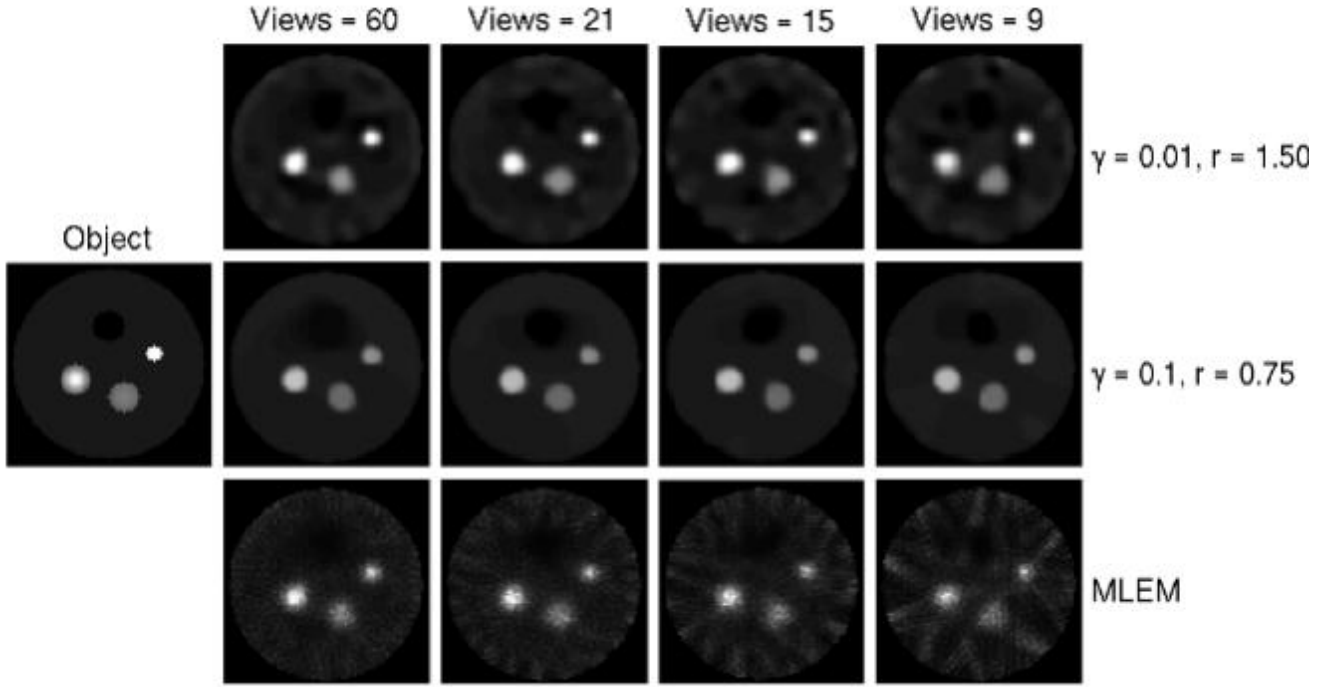

Figure 21. Reconstructions of GATE data simulated for 200s over different numbers of angles using the proposed algorithm and MLEM.

**Table 4.** Comparison of image quality metrics for images reconstructed from GATE data with the total scan time held constant as the number views decreased.

|  |  | 60 views | 21 views | 15 views | 9 views |
|---|---|---|---|---|---|
| $\gamma = 0.01, r = 1.50$ | **CC** | 0.946 | 0.945 | 0.942 | 0.946 |
|  | **SNR** | 17.48 | 20.57 | 18.31 | 6.72 |
|  | **FW10M** | 7 | 8 | 9 | 8 |
| $\gamma = 0.10, r = 0.75$ | **CC** | 0.942 | 0.940 | 0.941 | 0.945 |
|  | **SNR** | 21.83 | 24.55 | 20.95 | 81.18 |
|  | **FW10M** | 8 | 8 | 9 | 9 |
| MLEM | **CC** | 0.913 | 0.901 | 0.889 | 0.854 |
|  | **SNR** | 4.94 | 4.92 | 3.65 | 4.15 |
|  | **FW10M** | 8 | 11 | 9 | 8 |

*4.2.2. Constant scan time per view.* This set of simulations modeled a constant scan time per view (i.e., decreasing total scan time with decreasing number of views), representing the case where temporal sampling improves as the number of views decreases. Figures 22-24 present images reconstructed from 9 views with 10 seconds per view (compared to 66.67 seconds per view in figures 17b, 19 and 20).

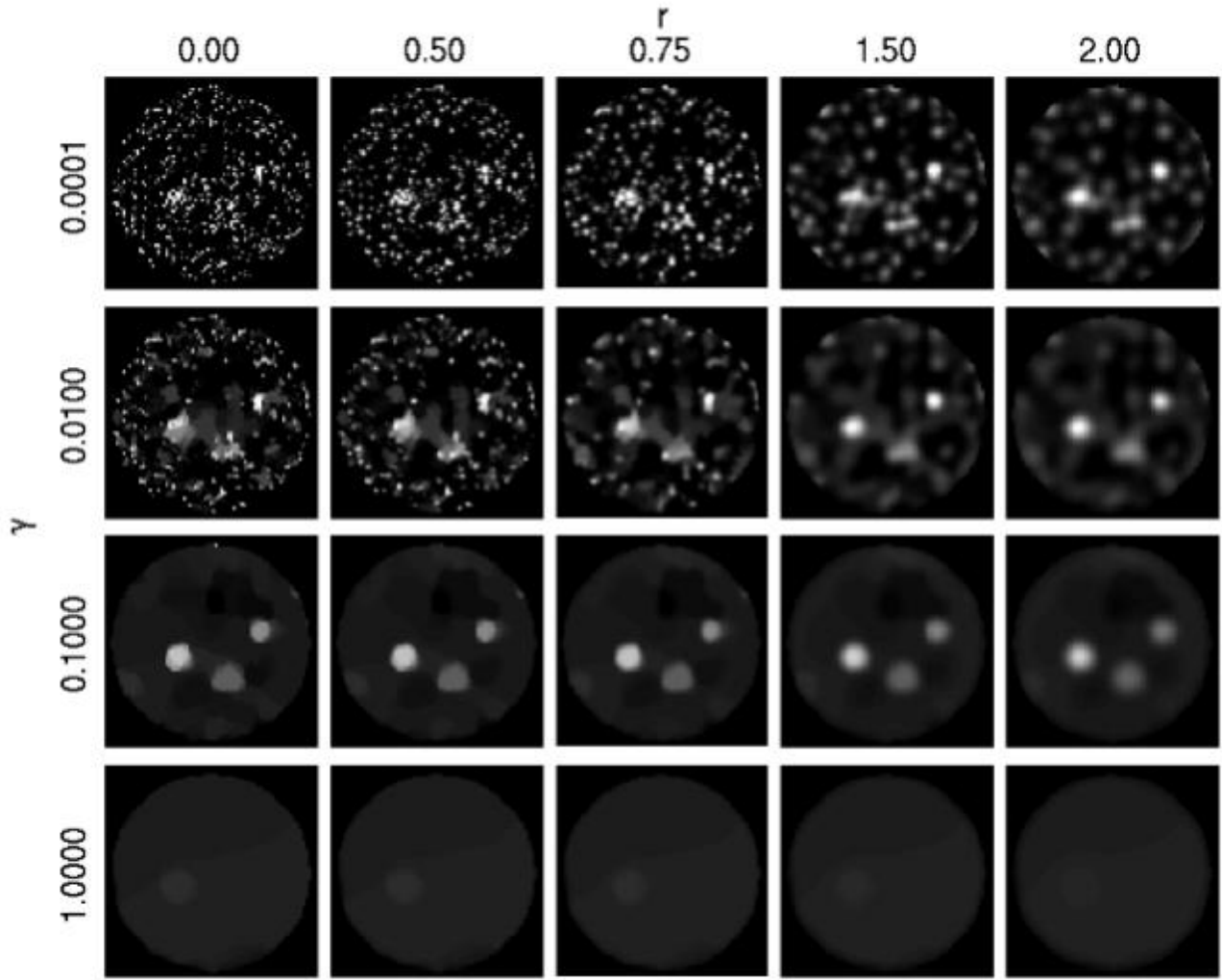

Figure 22.  Images reconstructed from 9 views of GATE data simulated for 30 seconds using the proposed algorithm with varying values of r and γ.

Images reconstructed from nine views using γ = 0.01 had lower reconstruction accuracy (CC < 0.9) compared to the images reconstructed from a 200 second scan time presented in the previous section. Using γ = 0.1, a maximum CC value of 0.921 occurred when $r$ = 0.75.  Similar to the 200 second scans, the CC varied by less than 2% across the range of $r$ values for γ = 0.1.  However, unlike the 200 second scans, the 30 second scans showed a larger variation in CC (~40%) across the range of $r$ values for γ not equal to 0.1. In addition to increasing CC, using γ = 0.1 resulted in reduced noise but increased blurring (higher FW10M) compared to γ = 0.01.

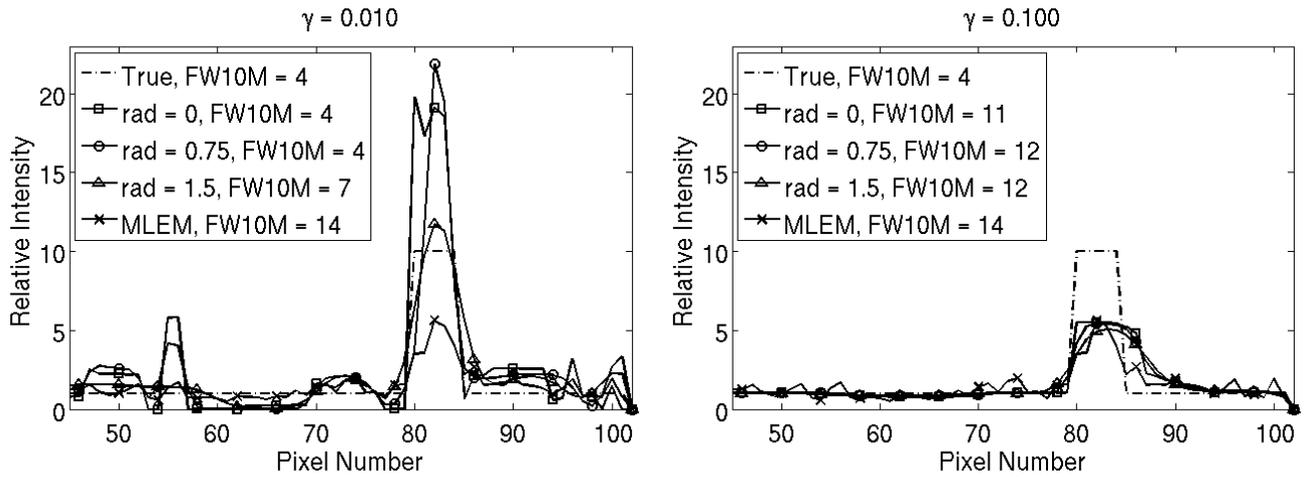

Figure 23. Central vertical profiles through images reconstructed from 9 views GATE data simulated for 30 seconds using the proposed algorithm with varying values of *r* and γ.

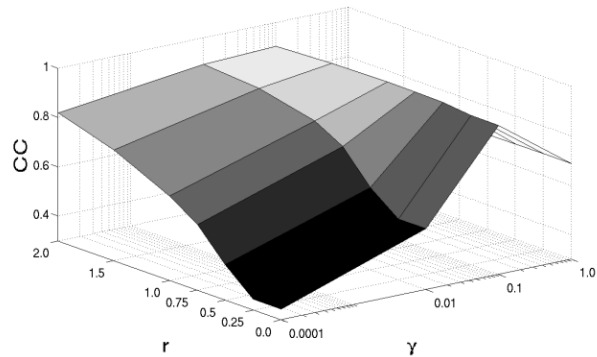

Figure 24. Plots depicting the CC over the range of studied *r* and γ parameters of images reconstructed from GATE data simulated for 9 views over 30 seconds.

Figure 25 compares images reconstructed with the proposed algorithm (γ = 0.01, *r* = 1.5 and γ = 0.01, *r* = 0.75) and MLEM from data acquired with a varying number of angular views (9 to 60) and a constant 10 second acquisition time for each angular position of the three-camera system. Thus the total scan time was 200, 70, 50, and 30 seconds for 60, 21, 15, and 9 views, respectively. Associated image quality metrics are presented in Table 5. As scan time and angular sampling decreased, images reconstructed using the proposed algorithm with γ = 0.01 and *r* = 1.50 show decreased accuracy compared to scans with less noise and the same angular sampling presented in the previous section. When reconstructing from 21 views, 15 views and 9 views, higher CC is achieved using γ = 0.1 and *r* = 0.75, compared to using γ = 0.01 and *r* = 1.50. In both cases, the proposed reconstruction algorithm provides higher CC and SNR than MLEM. For reconstructions using γ = 0.01 and *r* =

1.5, CC varied by 7.5% from 0.946 to 0.875 as the number of views was reduced from 60 to 9. A variation of only 2.6% from 0.942 to 0.921 was measured for images reconstructed using γ = 0.1 and $r$ = 0.75. These reductions were both less than the decrease measured for MLEM, which saw a decrease of 12.8%, from 0.915 to 0.798 as the number of views was reduced from 60 to 9.

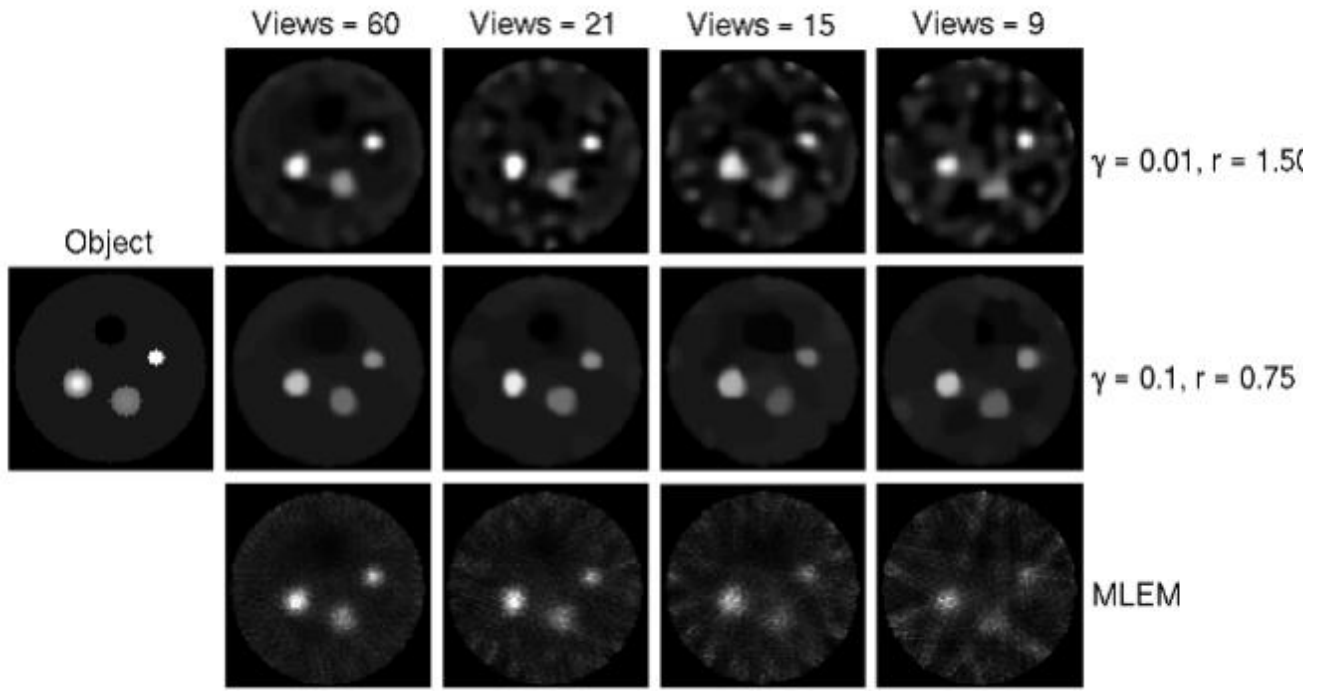

Figure 25. Images reconstructed using the proposed algorithm and MLEM from GATE data simulated with the same time per view for different numbers of views.

Table 5. Comparison of image quality metrics for images reconstructed from GATE data with varying number of views and constant scan time per view.

|  |  | 60 views | 21 views | 15 views | 9 views |
|---|---|---|---|---|---|
| γ = 0.01, $r$ = 1.50 | CC | 0.946 | 0.915 | 0.889 | 0.875 |
|  | SNR | 17.48 | 9.96 | 1.29 | 10.82 |
|  | FW10M | 7 | 6 | 9 | 7 |
| γ = 0.10, $r$ = 0.75 | CC | 0.942 | 0.937 | 0.908 | 0.921 |
|  | SNR | 21.83 | 5149.86 | 24.76 | 52.09 |
|  | FW10M | 8 | 8 | 10 | 12 |
| MLEM | CC | 0.915 | 0.885 | 0.829 | 0.798 |
|  | SNR | 4.94 | 2.98 | 3.49 | 4.06 |
|  | FW10M | 8 | 9 | 8 | 14 |

500    **5. Discussion**

The presented simulations investigated the proposed reconstruction technique over a range of objects, noise conditions, and angular sampling schemes. Overall, the results demonstrate that blurring and noise regularization increased with increasing values of $r$, the standard deviation of the Gaussian blurring kernel, and $\gamma$, the TV weighting parameter. For example, in the high-view case with noiseless data generated from the

505    system model, reconstructions using the lowest $\gamma$ value studied ($\gamma = 0.0001$) yielded the most accurate images for a given value of $r$. When the data were made inconsistent by the addition of Poisson noise, the optimal studied $\gamma$ value increased to $\gamma = 0.01$. These cases, in which data generated using the system matrix and the blurring model were used in reconstruction, indicate that accurate reconstruction is possible when the incorrect blurring model is used, as there was only a 2.5% decrease in the CC metric over all $r$ studied when $\gamma = 0.1$ and $\gamma = 0.01$.

510    However, in the few-view case, using an $r$ larger than $r_{true}$ caused the number of meaningful coefficients in the intermediate image $f$ to increase rapidly compared to using lower values of $r$. This indicates a less sparse image, limiting the effectiveness of exploiting gradient-magnitude sparsity to reduce the number of views needed for reconstruction. When an approximately accurate blurring model is used, the intermediate image $f$ is the most sparse in the gradient-magnitude sense. This may allow a greater reduction in the sampling necessary for

515    reconstruction.

When noisy data were generated using GATE Monte Carlo simulations, larger $r$ had a benefit when lower $\gamma$ were used. For instance, in the 9 view case when data were simulated for 200 seconds and images were reconstructed with $\gamma = 0.01$, the CC varied by 11% over the range of studied $r$ values, with a high $r$ value ($r = 1.5$) yielding the most accurate reconstructions. When $\gamma = 0.1$ was used, a lower $r$ ($r = 0.75$) yielded the most

520    accurate reconstructions. Similarly, when the scan time was decreased in the few-view case, $r = 2.0$ yielded the most accurate reconstructions when $\gamma = 0.01$ was used; however, the highest overall CC in the few-view, decreasing scan time case was obtained with $\gamma = 0.1$ and $r = 0.75$. Reconstructions using both $\gamma = 0.01$, $r = 2.0$ and γ = 0.1, $r$ = 0.75 have similar CC but different qualitative attributes (figures 21 and 25). The preferred parameter combination requires further study with observers. Overall, reconstructions from data generated using GATE simulations suggest that when the true blurring model is unknown and noise is present, lower values of γ (γ = 0.01 in this particular study) benefit from larger r values, while γ = 0.1 benefits from lower r values, with a smaller dependence on $r$. Since the inverse crime study demonstrated that smaller $r$ values result in a more sparse intermediate image, the combination of γ = 0.1 and $r$ = 0.75 may be advantageous for reconstruction from few-views.

The results also suggest that, when an appropriate value of the TV penalty term is included in the proposed reconstruction algorithm (γ = 0.01 or 0.1 for the cases studied), streaking artifacts are reduced compared to MLEM reconstructions. While images reconstructed with the proposed algorithm contain higher SNR, low frequency variations (patchy artifacts) were seen in high-noise simulation cases (figures 22 and 25). Low frequency, patchy artifacts have been noted in CT TV reconstructions from noisy data, and future work is required to quantify the impact of these artifacts on the ability of observers to identify objects of diagnostic interest. (Tang *et al* 2009).

The presented work suggests potential benefits of the proposed reconstruction algorithm compared to MLEM, however, additional work is required for a systematic comparison, including experimental investigation. One limitation of the presented work is that the simulations modeled 2D objects and acquisition, whereas SPECT data are acquired in three dimensions. We hypothesize that the principles and model presented in this work can be generalized to a 3D case with the expansion of the system matrix and applying the blurring-masking function in three dimensions. Additional studies are necessary to investigate this hypothesis. Reconstruction from multi-pinhole systems could be accomplished by modifying the system matrix to include contributions from all pinholes. Future work is also planned to apply the reconstruction technique to *in vivo* data to investigate the assumption that SPECT objects may be modeled as blurred piecewise constant objects. Future work will also investigate the performance of this algorithm for dynamic imaging from few-views.

## 6. Conclusions

This study proposed and characterized a sparsity-exploiting reconstruction algorithm for SPECT that is intended for few-view imaging and that phenomenologically models the object as piecewise constant subject to a blurring operation. While the reconstruction technique assumes a specific blurring model, the results demonstrate that the knowledge of the true blurring parameter is not required for accurate reconstruction, as the reconstruction algorithm has limited sensitivity to $r$ in the low noise cases and benefits from increasing $r$ in the high noise case. However, the results suggest that accurately modeling the blurring parameter provides increased gradient-magnitude sparsity, which may enable further reductions in sampling. The reconstructed images demonstrate that the reconstruction algorithm introduces low-frequency artifacts in the presence of noise, but eliminates streak artifacts due to angular undersampling. The effects of these artifacts on observers will be studied in future work. Overall, the results demonstrate preliminary feasibility of a sparsity-exploiting reconstruction algorithm which may be beneficial for few-view SPECT.

**Acknowledgments**

This work was supported in part by NIH R15 grant CA143713 and R01 grants CA158446, CA120540 and EB000225. The contents of this paper are solely the responsibility of the authors and do not necessarily represent the official views of the National Institutes of Health. This work is part of the project CSI: Computational Science in Imaging, supported by grant 274-07-0065 from the Danish Research Council for Technology and Production Sciences. The high performance computing resources used in this paper were funded by NSF grant OCI-0923037.